\documentclass[twocolumn,aps,amsmath,amssymb,prb]{revtex4}
\usepackage{graphicx}
\usepackage{bm}         
\usepackage{color}         
\usepackage{amsmath}
\usepackage{units}
\date{\today}
\begin{document}

\title{Topological transconductance quantization
in a four-terminal Josephson junction}

\author{Erik Eriksson, Roman-Pascal Riwar, Manuel Houzet, Julia S. Meyer}
\affiliation{Univ.~Grenoble Alpes, INAC-PHELIQS, F-38000 Grenoble, France}
\affiliation{CEA, INAC-PHELIQS, F-38000 Grenoble, France}
\author{Yuli~V.~Nazarov}
\affiliation{Kavli Institute of NanoScience, Delft University of Technology, Lorentzweg 1, NL-2628 CJ, Delft, The Netherlands}

\begin{abstract}
{ Recently we predicted that the Andreev bound state spectrum of 4-terminal Josephson junctions may possess topologically protected zero-energy Weyl singularities, which manifest themselves in a quantized transconductance in units of $4e^2/h$ when two of the terminals are voltage biased [R.-P. Riwar {\it et al.}, Nature Commun. {\bf 7}, 11167 (2016)]. Here, using the Landauer-B\"uttiker scattering theory, we compute numerically the currents flowing through such a structure in order to assess the conditions for observing this effect. We show that the voltage below which the transconductance becomes quantized is determined by the interplay of non-adiabatic transitions between Andreev bound states and inelastic relaxation processes. 
We demonstrate that the topological quantization of the transconductance can be observed at voltages of the order of $10^{-2} \Delta/e$, $\Delta$ being the superconducting gap in the leads.}
\end{abstract}

\maketitle

 \section{Introduction}

Topological phases of matter have attracted much interest in recent years.\cite{review-topo1,review-topo2} Starting with gapped phases such as topological insulators and superconductors, more recently gapless topological phases possessing topologically protected band crossings have been discovered.\cite{herring,weyl1,weyl2,weyl3} The topological properties of these systems are determined by their bandstructure and in particular the variation of the wavefunctions throughout the Brillouin zone.\cite{tknn,bernevig} Realizing topological phases is not an easy task and relies on finding the appropriate materials or combining different materials to engineer the required bandstructure.

Josephson junctions are a tool to probe topological properties~\cite{fu-kane,rokhinson,Badiane,erwann}, and they may possess interesting topological properties themselves.\cite{zhang-kane,vanHeck,padurariu,Riwar2016, yokoyama,giazotto,belzig} As some of us have pointed out recently~\cite{Riwar2016}, multiterminal Josephson junctions present an alternative to engineering topological materials. Josephson junctions host Andreev bound states (ABS) localized at the junction and whose energy is below the gap for the excitations in the leads. The spectrum of these ABS depends on the properties, both of the superconducting leads and the scattering region that connects them. The ABS energy is a function of the phase differences between the superconducting leads, which can be viewed as the quasi-momenta of the ABS \lq\lq bandstructure\rq\rq. This allows one to make an analogy between $n$-terminal junctions and $(n-1)$-dimensionals materials. We showed that the ABS spectrum of 4-terminal junctions made with conventional superconductors may possess Weyl singularities, corresponding to topologically protected zero-energy states. These Weyl singularities carry a topological charge $\pm1$. As a consequence the ABS pseudo-bandstructure as a function of two phase differences may possess a non-zero Chern number. We further showed that this non-zero Chern number leads to a quantized transconductance between two voltage-biased terminals.

The present work addresses the observability of this quantized transconductance in a transport experiment. The quantized transconductance is associated with adiabatic transport at fixed occupation of the ABS. On the other hand, a bias voltage is known to lead to multiple Andreev reflections~\cite{tinkham:1982,Averin1995,Bratus1995,Cuevas1996}, where a quasi-particle can be transferred from the occupied states below the superconducting gap to the empty states above the superconducting gap, leading to a dissipative current. At low bias, these processes may alternatively be described as resulting from Landau-Zener transitions between Andreev bound states.\cite{Averin1995} Here we compute the currents taking into account these processes in order to establish the voltage regime, where a quantized transconductance can be observed. { We find four different voltage regimes. At large voltages, the conductances are given by their normal state values. Decreasing the voltage, multiple Andreev reflections lead to a complicated voltage dependence with  pronounced subgap features. Further decreasing the voltage, a competition between Landau-Zener transitions and inelastic relaxation, which is modeled by an imaginary energy shift, takes place. Finally, at the lowest voltages, the dissipation vanishes and the transconductances reach their quantized values.}

The outline of the paper is the following. In Sec.~\ref{sec:2}, we provide the description of the Andreev spectrum and the topological properties of zero-energy states in a four-terminal Josephson junction through a time-reversal invariant normal region contacted to each superconducting terminal through a single channel. We illustrate the results with a random symmetric scattering matrix describing the normal-state properties of the junction (results for another one are given in the Appendix). In Sec.~\ref{sec:3}, we compute numerically the currents flowing through the voltage-biased junction within the scattering theory for out-of-equilibrium superconducting hybrid structures. In Sec.~\ref{sec:4}, we discuss the conditions for the observability of the transconductance quantization between two voltage-biased terminals. Our conclusions are given in Sec.~\ref{sec:5}.

\section{Topological characterization of the Andreev spectrum}  
\label{sec:2}
\subsection{Generalities}
\begin{figure}
\includegraphics[width=0.35\textwidth]{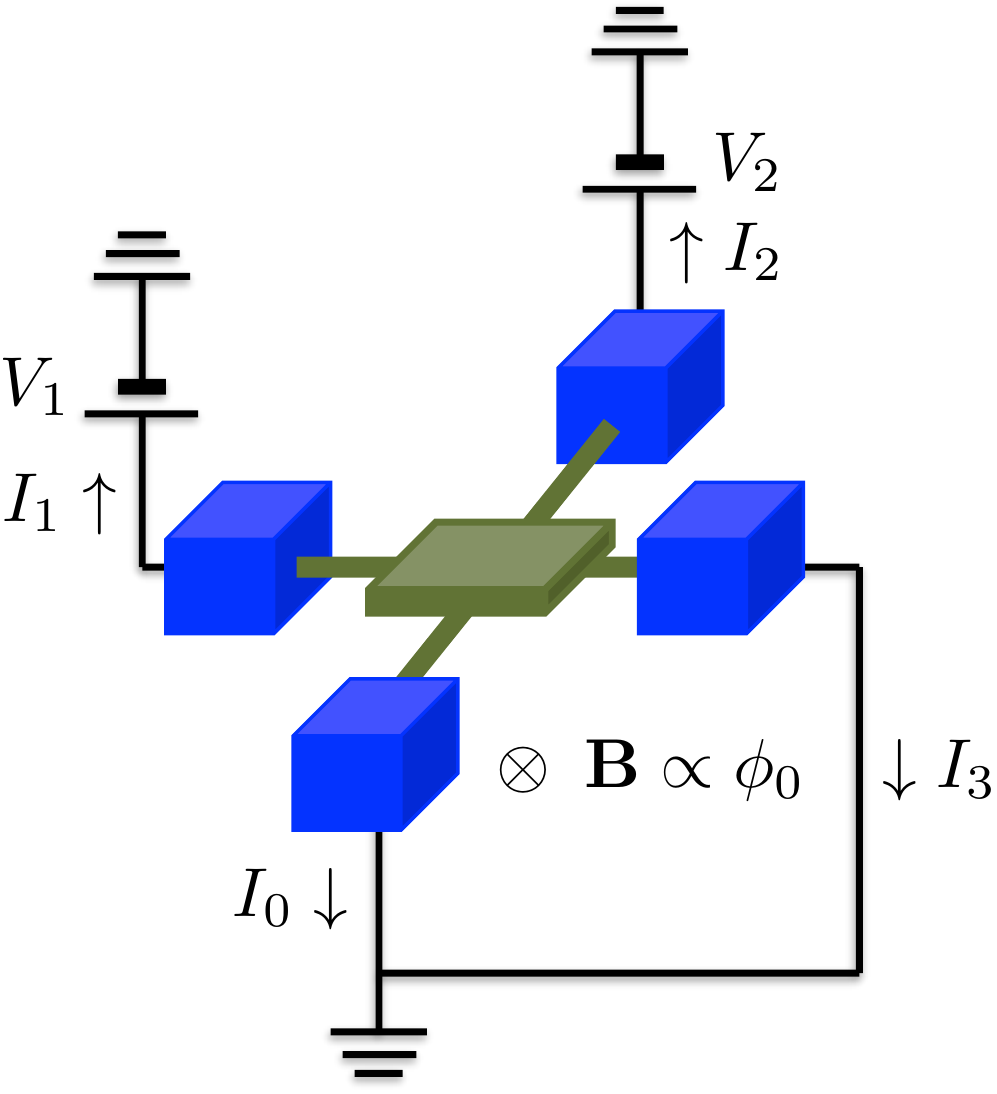}
\caption{\label{fig:setup} Sketch of a 4-terminal Josephson junction. The normal-state scattering region connecting the 4 leads is described by the scattering matrix $\hat{S}$. Leads $1$ and $2$ are voltage biased with voltages $V_1$ and $V_2$, respectively. The phase difference $\phi_0-\phi_3$ between lead $0$ and lead $3$ is controlled by a magnetic flux. Gauge invariance allows us to choose $\phi_3=0$. We compute the outgoing currents into all 4 leads.}
\end{figure}

We are considering a 4-terminal Josephson junction, the setup of which is shown in Fig.~\ref{fig:setup}. The superconducting leads are labeled by $\alpha=0,1,2,3$. The superconducting gap $\Delta$ is assumed to be the same for all leads. Their phases are given by $\phi_0,\phi_1,\phi_2,\phi_3$, where gauge invariance allows us to put $\phi_3=0$. The leads are connected through a short normal region characterized by an energy-independent (unitary) scattering matrix $\hat{S}$ for electrons. We further assume spin-rotation as well as time-reversal symmetry in the normal region, such that $\hat{S}$ is spin-independent and $\hat{S}=\hat{S}^T$. Thus $\hat S$ belongs to the circular orthogonal ensemble (COE). We further restrict to junctions having one transmitting channel per terminal, such that $\hat S$ is a $4\times4$ matrix.

The spectrum of ABS with energy $E$ ($|E|<\Delta$) in the junction is obtained by solving the eigenproblem\cite{Beenakker1991}
\begin{subequations}
\label{abseq}
\begin{eqnarray}
\psi_{e\alpha}(E)=\sum_\beta a(E)  S_{\alpha\beta} e^{i{\phi_\beta}} \psi_{h\beta}(E), \\
\psi_{h\alpha}(E)=\sum_\beta a(E)  S^*_{\alpha\beta} e^{-i{\phi_\beta}} \psi_{e\beta}(E). 
\end{eqnarray}
\end{subequations}
Here, $\psi_{e\alpha}$ and $\psi_{h\alpha}$ are electron and hole outgoing wavefunctions from the normal region to lead $\alpha$, respectively, and 
$a(E)=E/\Delta-i\sqrt{1-(E/\Delta)^2}$ is the Andreev reflection amplitude. 
Then, for each set of phases, Eq.~\eqref{abseq} admits for four solutions at energies $\pm E_1$ and $\pm E_2$ (with $0\leq E_1\leq E_2\leq\Delta$), which are pairwise opposite due to the built-in particle-hole symmetry in the theory of superconductivity. 

According to Ref.~\onlinecite{Riwar2016}, scattering matrices drawn out of the COE can admit for zero-energy Weyl points in the ABS spectrum. These Weyl points correspond to topologically protected crossings of the two solutions with energies $\pm E_1$ in the $(\phi_0,\phi_1,\phi_2)$-space of superconducting phases. Each crossing is characterized by a topological charge $Q=\pm 1$, where
\begin{equation}
Q=\frac1{2\pi}\int_{\cal S }d\bm{S}\cdot \bm{B}.
\end{equation} 
Here, $\cal S$ is a surface in the $(\phi_0,\phi_1,\phi_2)$-space that encloses the Weyl point, $d\bm{S}$ is an element of that surface, and 
\begin{equation}
\bm{B}\equiv (B_0,B_1,B_2)=i\,\sum_{\alpha=0}^3\bm{\partial}\psi_{e\alpha}^*\times \bm{\partial}\psi_{e\alpha},
\end{equation}
where $\bm{\partial}=(\partial_{\phi_0},\partial_{\phi_1},\partial_{\phi_2})$ is the Berry curvature associated with a normalized eigenstate with energy $-E_1$ (with $\sum_{\alpha}|\psi_{e\alpha}|^2=1$). 
Time-reversal symmetry, together with the fermion-doubling theorem,\cite{Nielsen1981} imposes that the Weyl points appear in groups of four: there are two Weyl points of a given charge at $\pm(\phi_0^{(1)},\phi_1^{(1)},\phi_2^{(1)})$, as well as two Weyl points of the opposite charge at $\pm(\phi_0^{(2)},\phi_1^{(2)},\phi_2^{(2)})$. For definiteness, we chose $0\leq\phi_0^{(1)}\leq \phi_0^{(2)}\leq \pi$. For phases $\phi_0\neq\phi_0^{(i)}$, the Andreev spectrum is gapped in the entire $(\phi_1,\phi_2)$-plane.

Subsequently, we define a (topological) Chern number in the $(\phi_1,\phi_2)$-plane,
\begin{equation}
C_{12}(\phi_0)=-C_{21}(\phi_0)=\frac1{2\pi}\int_{-\pi}^{\pi}\!\!\!d\phi_1\int_{-\pi}^{\pi}\!\!\!d\phi_2\, B_0(\phi_0,\phi_1,\phi_2).
\end{equation} 
At $\phi_0=0$, the setup is effectively a three-terminal junction, which does not admit for topologically protected crossings.\cite{Riwar2016} Time-reversal symmetry imposes that the Chern number $C_{12}(\phi_0=0)=0$. Increasing the phase $\phi_0$, the Chern number changes when crossing $\phi_0^{(i)}$ by the charge of the corresponding Weyl point. Thus, we deduce that $C_{12}(\phi_0)=0$ in the regions $0<\phi_0<\phi_0^{(1)}$ and $\phi_0^{(2)}<\phi_0<\pi$, while it takes a value $1$ or $-1$ in the intermediate region $\phi_0^{(1)}<\phi_0<\phi_0^{(2)}$. Furthermore, $C_{12}(-\phi_0)=-C_{12}(\phi_0)$. 

According to adiabatic perturbation theory~\cite{Riwar2016}, the Chern number determines the transconductance between two-voltage-biased terminals 1 and 2, $G_{12}=-(4e^2/h)C_{12}$, at sufficiently low voltage biases. To probe the transconductance quantization beyond the adiabatic regime, we will perform a numerical calculation of the current at arbitrary voltages for two specific setups (cf. Sec.~\ref{sec:3} and Appendix). Below we motivate our choice for the two different scattering matrices describing these setups.

\subsection{Examples}

To obtain systems with Weyl points, we generate random symmetric scattering matrices $\hat{S}$ within COE. This is done by first generating random hermitian matrices $H$ from the Gaussian unitary ensemble, and then forming $\hat{S} = U^T U$, where $U$ is the unitary matrix that diagonalizes $H$, i.e.,~$H= U^{\dagger} D U$ with $D$ a real diagonal matrix. Around 5\% of these matrices admit for Weyl points.
 
 To observe the  topological quantization of the transconductance, it is favorable to have a large gap in the Andreev spectrum between the Weyl points. 
 Thus, for each scattering matrix with Weyl points, we determine the largest possible gap in the $(\phi_1, \phi_2)$-plane for all $\phi_0$ in between the Weyl points,
\begin{equation}
{E_g=}\,\max_{\phi_0^{(1)}<\phi_0<\phi_0^{(2)}}\,\min_{\phi_1,\phi_2}E_1(\phi_0,\phi_1,\phi_2),
\end{equation} 
Furthermore, we do the same for all the remaining $\phi_0$ in the intervals $0<\phi_0<\phi_0^{(1)}$ and $\phi_0^{(2)}<\phi_0<\pi$.
A histogram of the smallest of these maximal gaps for an ensemble of topological scattering matrices is shown in { Appendix \ref{app-A}}. We do not find any gap larger than around $E = 0.12 \Delta$. For our simulation, we choose a topological scattering matrix with a gap close to that value.

\begin{figure}
\centering
\includegraphics[width=0.52\textwidth]{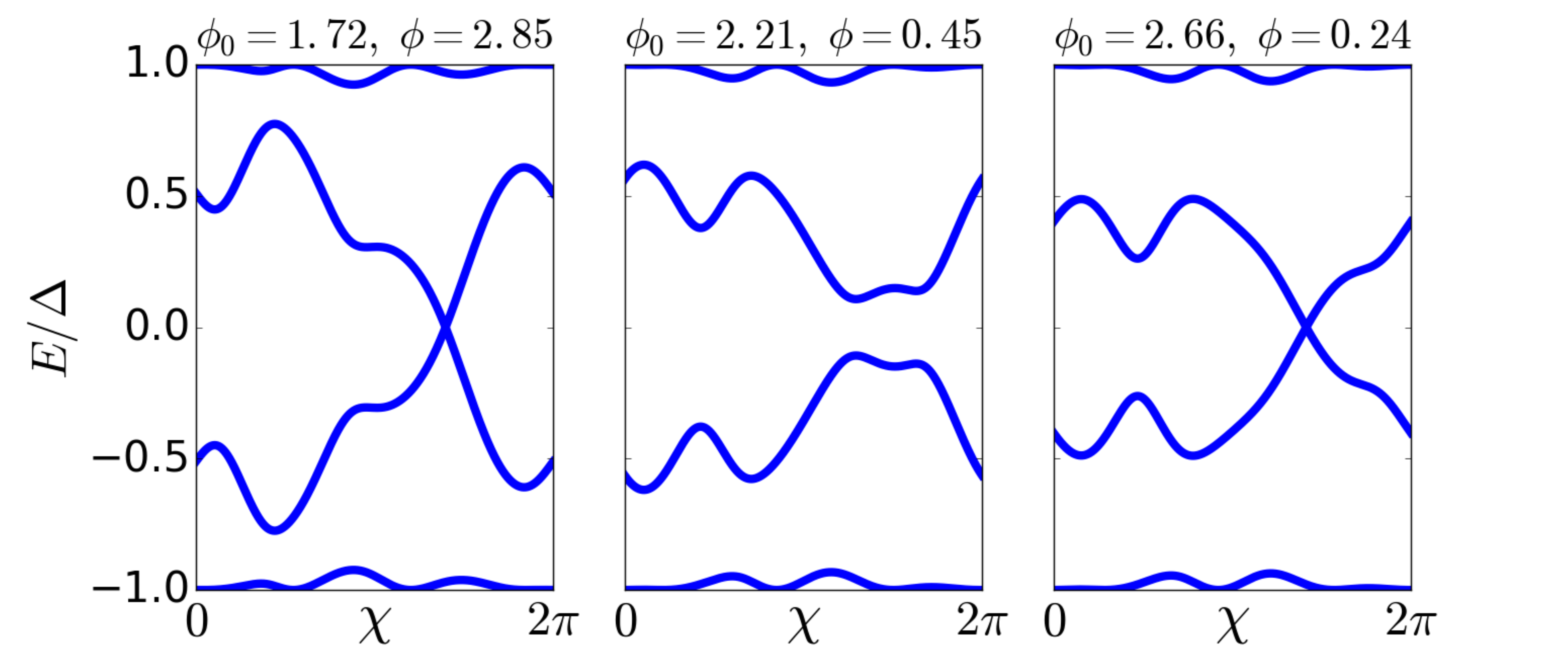}
\caption{Cuts through the Andreev bound state spectrum for the scattering matrix $\hat{S}_1$. From left to right: At phase $\phi_0 = 1.72$ (Weyl point), at phase $\phi_0 = 2.21$ where the gap is the largest within the topological region ($E=0.11\Delta$), and at phase $\phi_0 =2.66$ (Weyl point). The spectra are along the lines $(\phi_1=\chi , \phi_2 = 3\chi+\phi)$, where the phase shift $\phi$ is chosen such that the cut goes through the point in the $(\phi_1,\phi_2)$-plane with the smallest gap.}
\label{fig:ABS}
\end{figure}

Specifically, we use the matrix
\begin{widetext}
\begin{equation} \label{S1}
\hat{S}_1 = \left(  \begin{matrix}
  0.299+i0.091 & {-0.547}-i0.171 & -0.190-i0.474 &   -0.543-i0.140 \\
  {-0.547}-i0.171 &  0.271+i0.306 & -0.334-i0.182 &  -0.288-i0.527\\
  -0.190-i0.474 & -0.334-i0.182  & 0.348+i0.565 & -0.369-i0.140 \\
  -0.543-i0.140 & -0.288-i0.527 & -0.369-i0.140 &
   0.317+i0.263
 \end{matrix}  \right),
\end{equation}
\end{widetext}
which has four Weyl points at $(\phi_0,\phi_1,\phi_2) = \pm (1.72,-1.89,-2.82)$, with charge $-1$, and $(\phi_0,\phi_1,\phi_2) = \pm (2.66,-1.84,1.01)$, with charge $+1$. Taking $\phi_0$ as a control parameter, its maximal gap in the $(\phi_1,\phi_2)$-plane for $\phi_0$ in between the two Weyl points is $E=0.11\Delta$ and is realized at $\phi_0 = \pm2.21$. 
See Fig.~\ref{fig:ABS} for some examples of the ABS spectrum of the four-terminal setup.

In the Appendix, we show results for a second scattering matrix, which has a smaller gap in between its Weyl points.

\section{Current-voltage characteristics}
\label{sec:3}

In this section we use the Landauer-B\"uttiker scattering formalism extended to superconducting hybrid structures to calculate the currents flowing through the setup at arbitrary voltage biases.\cite{Averin1995} We compare the numerical results with the prediction of the transconductance quantization at sufficiently low voltages.

\subsection{Formalism}

To obtain the transconductances $G_{12}$ and $G_{21}$, we need to voltage bias leads 1 and 2. We will consider that they are voltage biased with commensurate voltages $V_1= n_1 V$ and $V_2= n_2 V$ ($n_1,n_2$ integers), while $V_0=V_3=0$. The d.c.~currents flowing to the leads also depend on the phase bias $\phi_0$, as well as on a phase shift $\phi$ between the time-dependent phases, $\phi_1(t)=2eV_1t$ and $\phi_2(t)=2eV_2t+\phi$ {(with $e>0$)}. They are given by {(below we set $\hbar=k_B=1$, unless they are explicitly written out)}
\begin{widetext}
\begin{equation} 
\label{current}
I_{\alpha} = I^N_{\alpha} - \frac{e}{2\pi} \int\mathrm{d}E \, J^2(E)  \tanh\frac{E}{2T} 
\left\{ 
	2 \mathrm{Re} [a(E) \psi^{\alpha E}_{h, \alpha} (E)] 
	+ 
	\sum_{\beta,k} 
	 ( |a_k(E) |^2+1) 
		\left(
			|\psi^{\beta E}_{h \alpha} (E+k{e}V)|^2
			-|\psi^{\beta E}_{e\alpha} (E+k{e}V)|^2
		\right) 
\right\},
\end{equation}
where $I^N_{\alpha} = (2e^2/h) \sum_{\beta} |S_{\alpha \beta} |^2 (V_{\beta} - V_{\alpha})$ is the normal-state current, $T$ is the temperature, $J(E)=\sqrt{1-|a(E)|^2}$, and $a_k(E)=a(E+keV)$. Here $a(E)=[E+i\Gamma-i\sqrt{\Delta^2-(E+i\Gamma)^2}]/\Delta$ generalizes the Andreev reflection amplitude to energies below and above the gap. It includes a phenomenological broadening parameter $\Gamma$, also known as Dynes parameter (see below).\cite{dynes} The outgoing electron and hole wavefunctions in lead $\alpha$ associated with an incoming electron-like state from lead $\beta$ and with energy $E$ are given by the set of equations
\begin{subequations}
\label{MAR}
\begin{eqnarray}
\psi^{\beta E}_{e\alpha}(E+keV) &=& \sum_{\gamma} e^{i(\phi_{\alpha} - \phi_{\gamma})} {S}_{\alpha\gamma} \left[a_k(E - n_{\gamma}eV +n_{\alpha}eV) \psi^{\beta E}_{h\gamma}(E+keV- n_{\gamma}eV +n_{\alpha}eV)+\delta_{\beta,\gamma} {\delta_{k,n_\gamma-n_\alpha}}\right],
\\
\psi^{\beta E}_{h\alpha}(E+keV) &=& \sum_{\gamma} e^{i(-\phi_{\alpha} + \phi_{\gamma})} {S}^*_{\alpha\gamma} a_k(E+ n_{\gamma}eV -n_{\alpha}eV) \psi^{\beta E}_{e\gamma}(E+keV+ n_{\gamma}eV -n_{\alpha}eV),
\end{eqnarray}
\end{subequations}
\end{widetext}
which take into account inelastic scattering processes due to voltage biases. Here $n_0=n_3=0$, $\phi_1=\phi_3=0$, and $\phi_2=\phi$.

The Dynes parameter generates a finite density of states at all energies below $\Delta$,
$\nu(E)=\nu_0\mathrm{Re}\{[1+a^2(E)]/[1-a^2(E)]\}$, where $\nu_0$ is the density of states in the normal state. In particular, $\nu(E\ll\Delta)\simeq \nu_0\Gamma/\Delta$ at $\Gamma\ll\Delta$. Thus it admits for inelastic relaxation of subgap states within the junction by coupling them with the small density of states in the leads. By contrast, when $\Gamma=0$, quasiparticles can only relax their energy after performing successive Andreev reflections in the subgap region until they reach $\Delta$.

Note that topological conductance quantization has been predicted for incommensurate voltages.\cite{Riwar2016} As we work with commensurate voltages, the currents contain a Josephson-like contribution that depends periodically on the phase shift $\phi$ between the voltage-biased leads. This contribution has a small amplitude for commensurability ratios $n_1/n_2$ different from 1 (see Fig.~\ref{fig:offset} in the Appendix) and would vanish for incommensurate voltage biases. To extract the $\phi$-independent part of the currents, we perform an average of the currents \eqref{current} over the phase shift $\phi$.

The solution of the coupled equations \eqref{MAR} is implemented numerically as a matrix equation problem, making use of \verb|python|'s \verb|scipy.linalg| library (the \verb|solve_banded| algorithm for solving a matrix equation with a sparse banded matrix). From the obtained wavefunctions, the currents are computed by the integration over energy in Eq.~(\ref{current}). For the integration we use direct summation over energies $-3\Delta \leq E \leq 3\Delta$, with a sampling distance d$E = 0.2 eV$. The computation time scales as $\sim V^{-2}$. At voltage $V=0.007 \Delta/e$ and fixed $\phi_0$, computing all the currents for a single phase shift $\phi$ takes around 4h on a single CPU.  Although the time can be reduced by parallelizing the integration over energy in Eq.~(\ref{current}), and the averaging over phase shifts (in practice, 10 equidistant phase shifts were sufficient to perform that average), this still limits the voltage range that we are able to efficiently probe. 

In the next subsection, we show the numerical results for the scattering matrix $\hat S_1$, Eq.~\eqref{S1}. Results for a different scattering matrix $\hat S_2$ are shown in the Appendix.

\subsection{Results}

In Fig.~\ref{fig:currents} we show the $I-V$ curves for the 4-terminal setup with scattering matrix $\hat{S}_1$ 
and a Dynes parameter $\Gamma = 0.002\Delta$. To extract the transconductances, we use two sets of values $(n_1,n_2)$.
Shown in the figure are the $I-V$ curves using $(n_1=2,n_2=3)$ and two different values of the phase bias $\phi_0$. Note that the currents $I_1$ and $I_2$ have to tend to zero as voltage tends to zero. By contrast, a Josephson current may circulate in the ring between leads $0$ and $3$ at non-zero $\phi_0$. This is seen in the top panel of Fig.~\ref{fig:currents}, where $I_3$ tends to $-I_0\neq0$ as the voltage tends to zero.

\begin{figure}[t!]
    \includegraphics[width=0.5\textwidth]{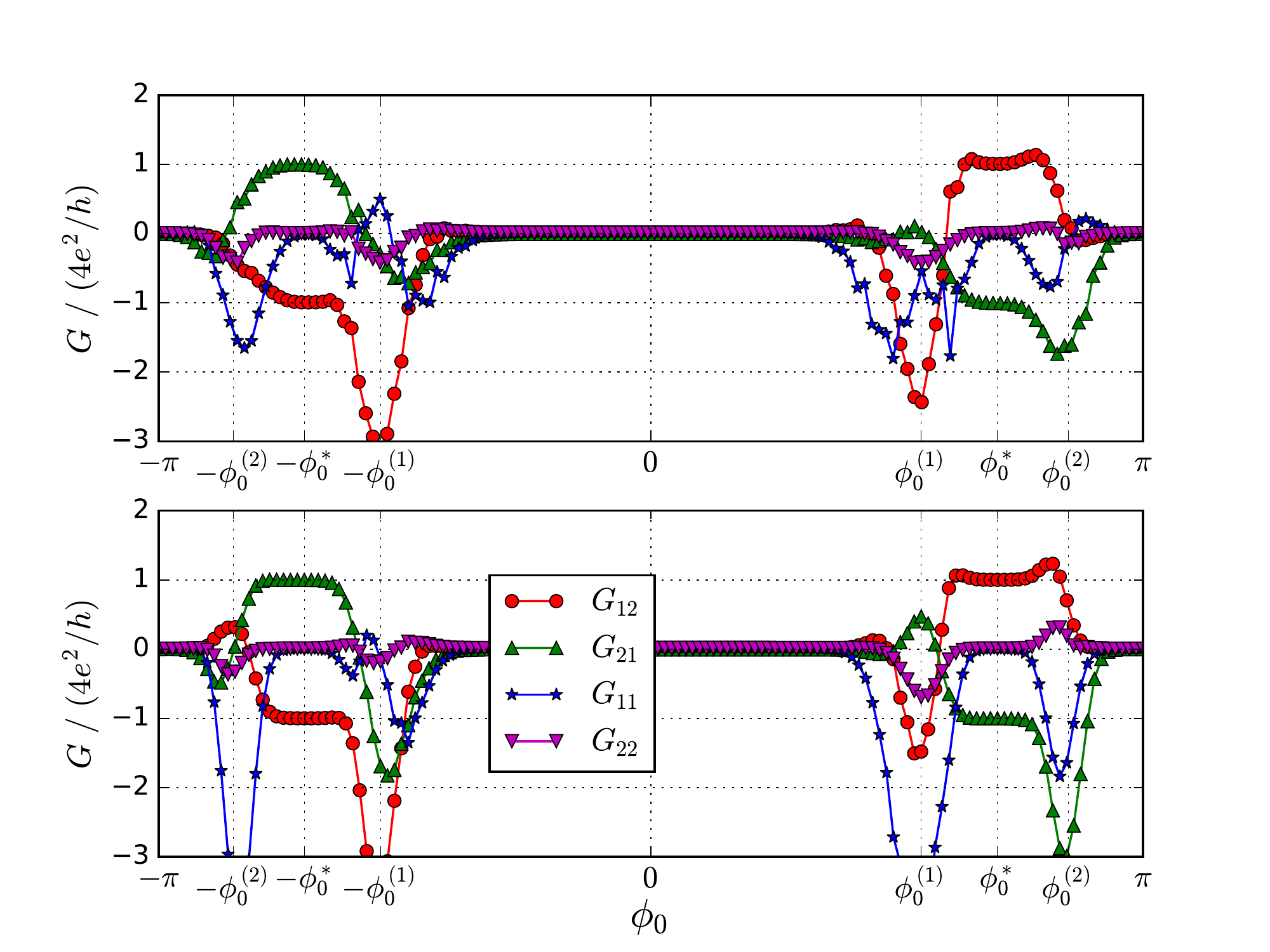}
    \caption{Conductances as a function of phase $\phi_0$ at fixed voltage.  Top: $V=0.005 \Delta/e$. Bottom: $V=0.003 \Delta/e$. The quantized conductance plateaus are clearly visible. Around the topological transitions at $\pm \phi_0^{(1)}=\pm 1.72$ and $\pm \phi_0^{(2)}=\pm 2.66$, conductance quantization breaks down because the gap closes and dissipation becomes large. 
      } \label{fig:G-phi0}
\end{figure}

{ From the computed currents $I_{\alpha}$ for the two sets of voltage biases, we obtain the conductance matrix $\hat G$, defined as $I_{\alpha} = \sum_\beta G_{\alpha \beta} V_{\beta}$, for $\phi_0=0$ (trivial region) and $\phi_0=2.21$ (topological region). Its elements as a function of voltage $V$ are shown in Fig.~\ref{fig:conductance}.  At very low voltages, the direct conductances vanish, while the transconductances become quantized, $G_{\alpha \beta} = -(4e^2/h) C_{\alpha \beta}$. We now fix the voltage to a value small enough to observe conductance quantization and vary the control parameter $\phi_0$.
In Fig.~\ref{fig:G-phi0}, the dependence of the transconductance as a function of $\phi_0$ is shown for two different voltages. We see that the transconductance quantization holds for $\phi_0$ not too close to the values $\pm\phi_0^{(1)},\pm \phi_0^{(2)}$, where the topological transitions take place. }

 \begin{figure}
	\centering
	    \includegraphics[width=0.45\textwidth]{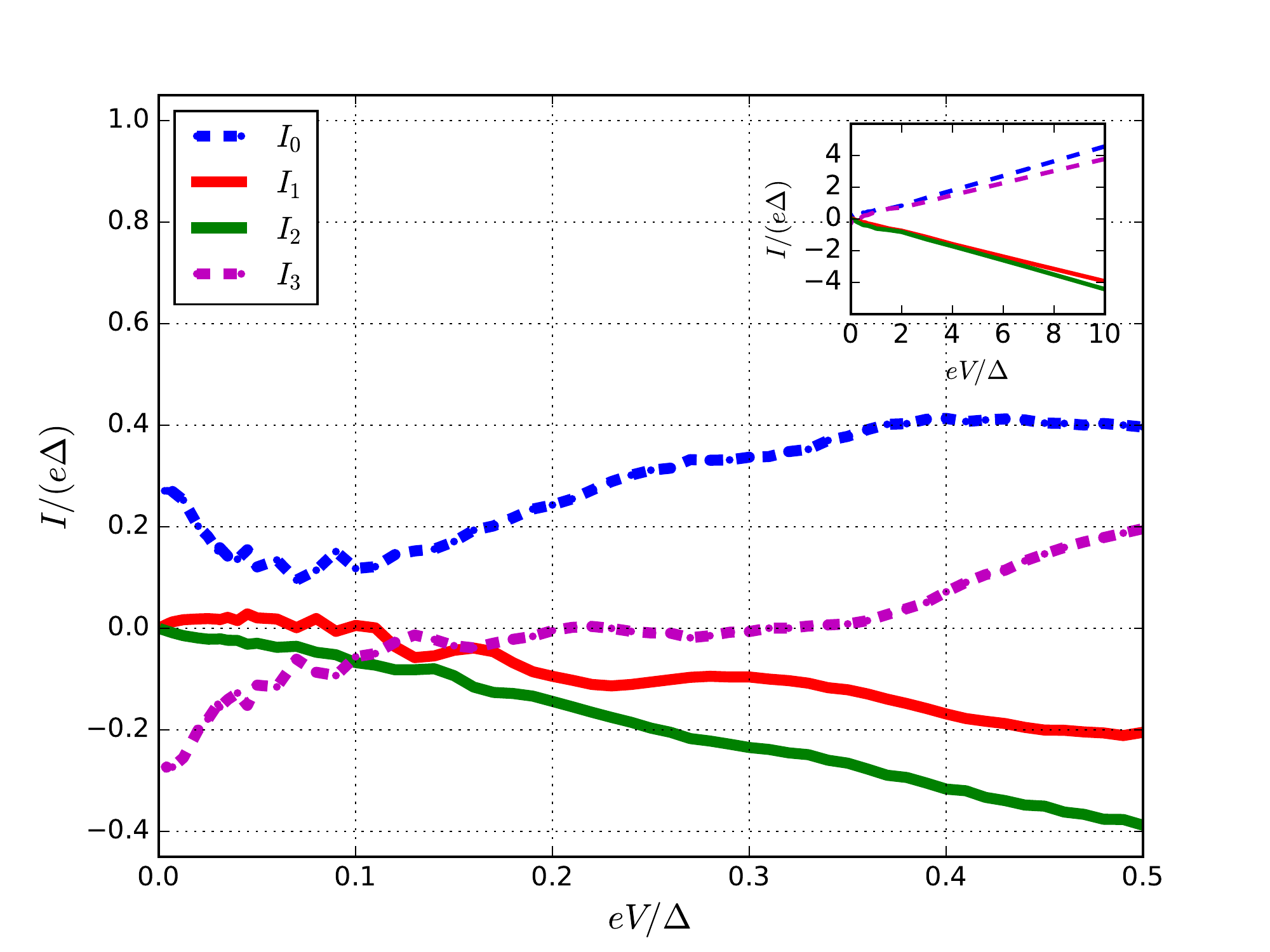}\\
	    \includegraphics[width=0.45\textwidth]{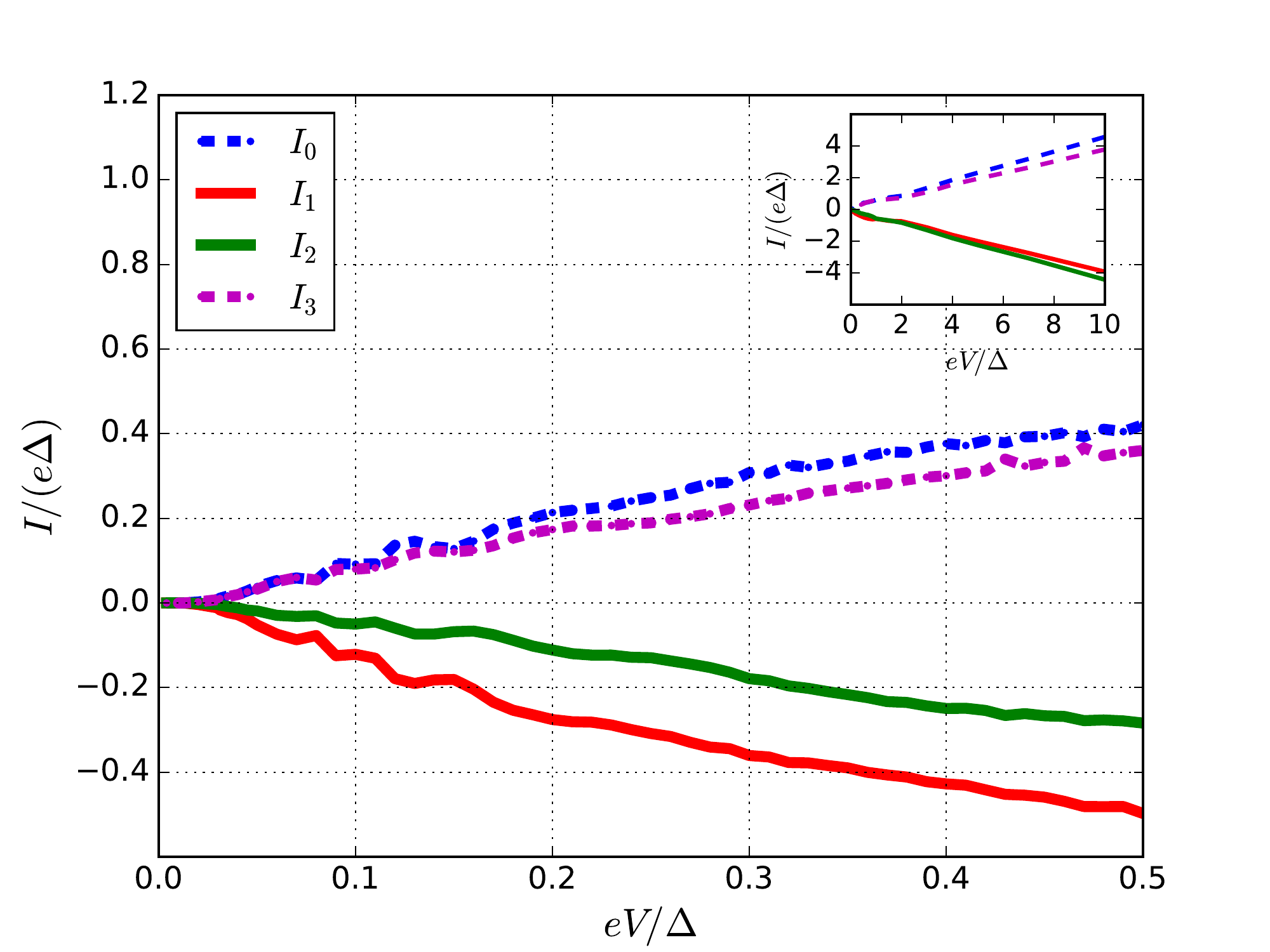}
	\caption{The currents $I_0,I_1,I_2,I_3$ as function of voltage for the scattering matrix $\hat{S}_1$. The voltages in terminals $1$ and $2$ are given as $V_1 = 2V$ and $V_2 =3V$, respectively. The Dynes parameter is set to $\Gamma=0.002\Delta$. Top: At phase $\phi_0 = 2.21$ in the topological region. Bottom: At phase $\phi_0=0$ in the trivial region. We have used an average over $N=10$ phase shifts $\phi$. The insets show a larger range of voltages.}
	\label{fig:currents}
	\end{figure}

  \begin{figure}
	\centering
		\includegraphics[width=0.49\textwidth]{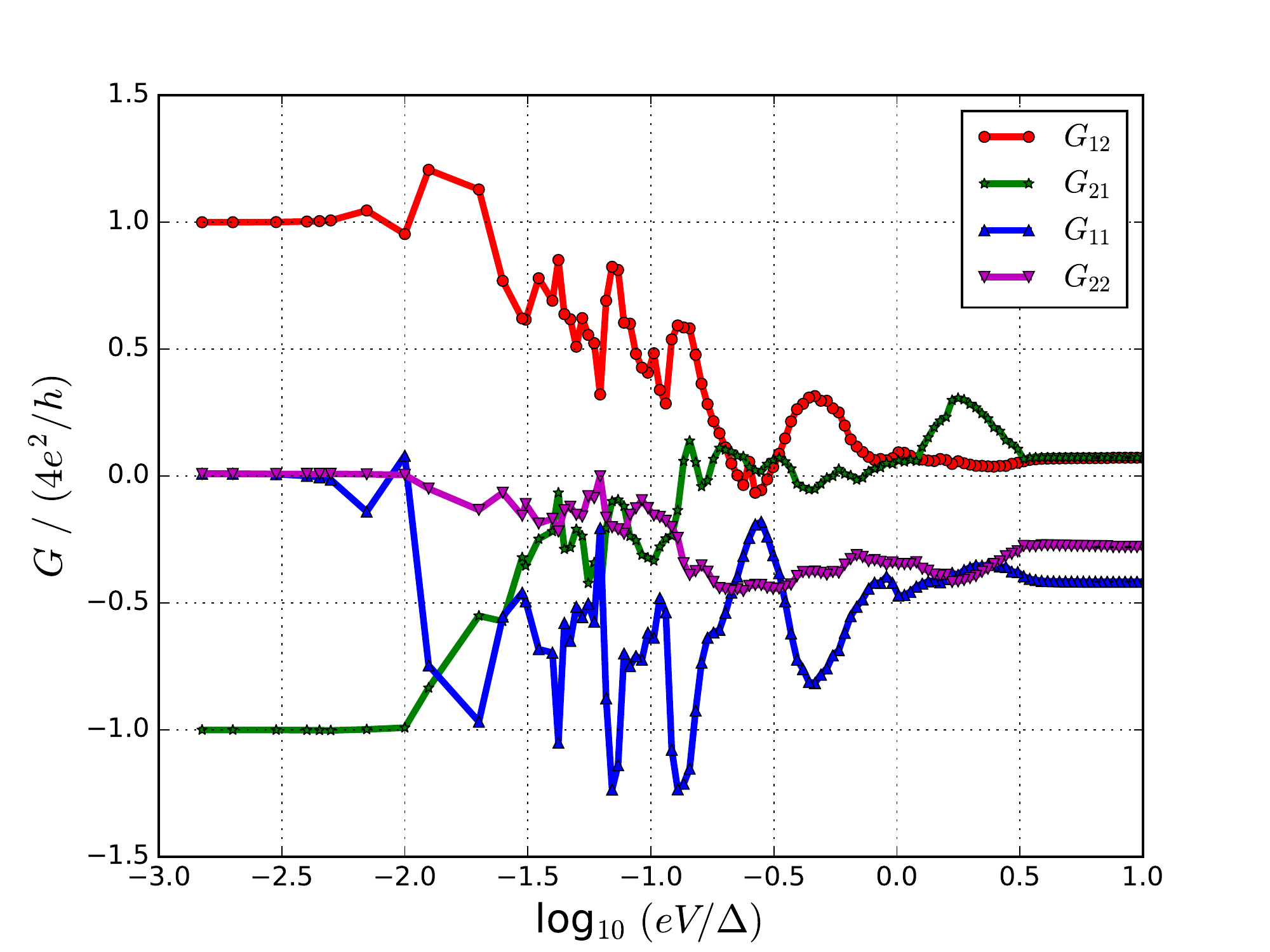}\\
		\includegraphics[width=0.49\textwidth]{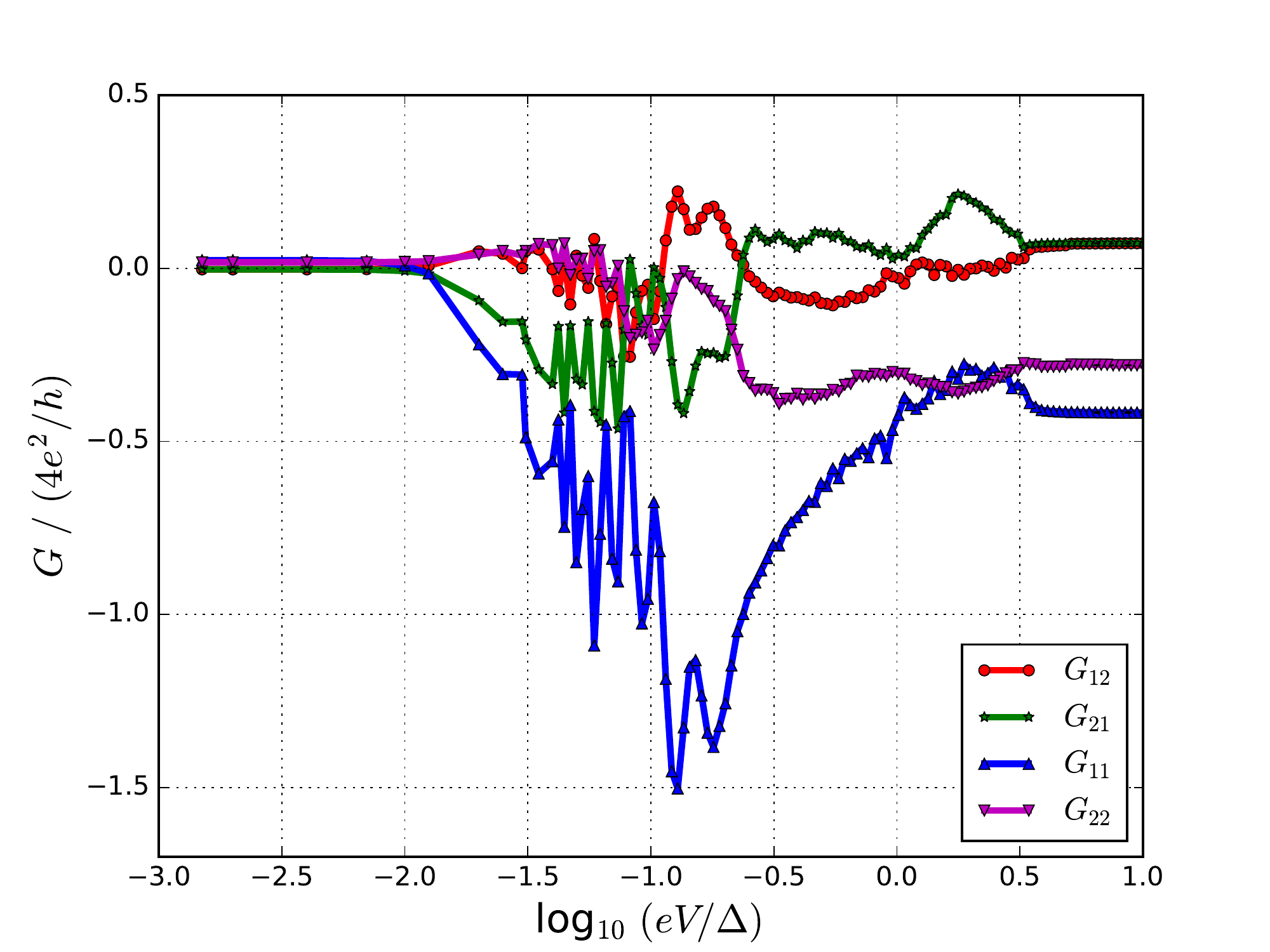}
	\caption{The conductances $G_{12},G_{21},G_{11}$, and $G_{22}$ between the voltage-biased leads $1,2$ as a function of voltage in logarithmic scale. The conductances are obtained from the currents shown in Fig.~\ref{fig:currents} and a similar set obtained with a different voltage ratio, $V_1 = V$ and $V_2 =3V$. Top: At phase $\phi_0 = 2.21$ in the topological region. Bottom: At phase $\phi_0=0$ in the trivial region. The expected quantization of the transconductance is seen for voltages $eV/\Delta\lesssim 0.01$.}
	\label{fig:conductance}
\end{figure}

\begin{figure}
	\centering
		\includegraphics[width=0.47\textwidth]{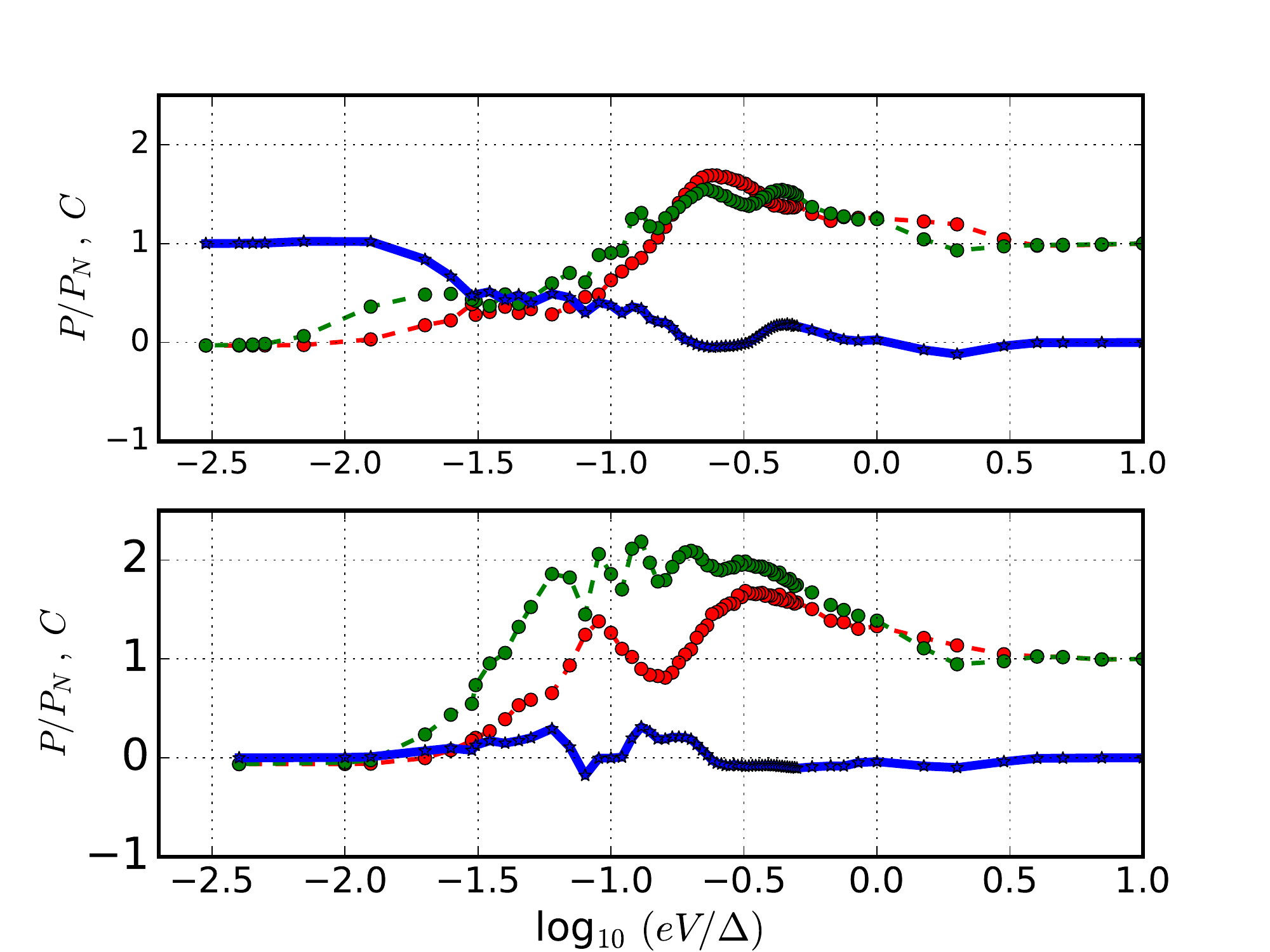}
	\caption{Dissipation $P/P_N$ (dashed lines) and chirality $C$ (solid lines) as a function of voltage in logarithmic scale. Green curves correspond to $V_1=2V,V_2=3V$ and red curves to $V_1=V,V_2=3V$.  Top: At phase $\phi_0 = 2.21$ in the topological region. The chirality tends to 1 as the dissipation tends to zero at low enough voltages. Bottom: At phase $\phi_0=0$ in the trivial region. }
	\label{fig:dissipation}
\end{figure}

To help interpreting the results, we compute the dissipation of the system as a function of voltage, defined as $P \equiv \sum_{\alpha} I_{\alpha}  V_{\alpha}$. Using the two sets of voltages, we also compute the chirality, defined as 
\begin{equation} \label{chirality}
C \equiv {\left( \frac{h}{8 e^2}\right)}   \frac{I'_1 V_1 - I_1V'_1 + I'_2 V_2 - I_2 V'_2 }{V_1 V'_2 - V_2 V'_1}
\end{equation}
when $V_0=V_3=0$. Here the primed variables are computed using the set $n_1=2,n_2=3$ and the unprimed using $n_1=1,n_2=3$. {The chirality selects the antisymmetric part of the conductance matrix. In a linear response regime, it reduces to $C=(h/8e^2)(G_{12}-G_{21})$.  In particular, in the presence of time-reversal symmetry, it vanishes in the normal state.} In Fig.~\ref{fig:dissipation}, we plot the chirality $C$ as a function of voltage for the same structure as in Figs.~\ref{fig:currents}-\ref{fig:conductance}, together with the normalized dissipation $P/P_N$, where $P_N \equiv  \sum_{\alpha} I^N_{\alpha}  V_{\alpha}$ is the normal-state dissipation.

\section{Discussion}
\label{sec:4}

{The adiabatic perturbation theory that was used in Ref.~\onlinecite{Riwar2016} to predict the transconductance quantization requires the Andreev levels to retain their equilibrium occupations. In that regime, direct conductances vanish. On the other hand, multiple Andreev reflections allow for quasiparticle transfer between the leads by overcoming the energy gap $2\Delta$. Thus, they result in charge transport at subgap voltages. At low voltages, these multiple Andreev reflections can be related with non-adiabatic transitions of quasiparticles occupying different branches of the Andreev spectrum as the phases increase linearly with time due to the voltage biases. Therefore the two regimes described above are competing. 
We expect that the transconductance quantization holds provided that an inelastic scattering process restores equilibrium occupation of the subgap states while suppressing multiple Andreev reflections.  The Dynes parameter $\Gamma$ provides such a mechanism, while essentially preserving the superconducting gap if $\Gamma\ll\Delta$. 

{ In Fig.~\ref{fig:conductance} showing the conductance as a function of voltage, we can distinguish four different voltage regimes.

We see that, for high voltages $V \gg \Delta/e$, the conductance matrix elements match their normal-state values, $G^N_{\alpha\beta}=(2e^2/h)(|S_{\alpha\beta}|^2-\delta_{\alpha\beta})$. (Note that $G^N_{\alpha\alpha}<0$ and $G^N_{\alpha\neq\beta}>0$ due to the chosen conventions for the current directions.) 
In particular, $G^N_{11}=-0.42\times(4e^2/h)$, $G^N_{22}=-0.28\times(4e^2/h)$, and $G^N_{12}=G^N_{21}=0.07\times(4e^2/h)$, {corresponding to $C=0$ (cf.~Figs.~\ref{fig:conductance} and \ref{fig:dissipation})}.

At lower voltages, $0.1 \Delta/e\lesssim V \lesssim 2 \Delta/e$, we observe a complex dependence of the direct conductances as well as the transconductances, with resonant features that are related with multiple Andreev reflections involving various leads.\cite{cuevas-pothier,samuelsson-houzet,quartets1,quartets2,physica,quartets3} 

At even lower voltages, $V \lesssim 0.1\Delta/e$, the interplay between Landau-Zener transitions and inelastic relaxation becomes important. If $\Gamma$ is larger than the Landau-Zener transition rate between the states with energy $-E_1$ and $E_1$, it restores the equilibrium occupations, where the state with energy $-E_1$ is occupied and the state with $E_1$ is empty, throughout most of the time evolution. 

Thus, at very low voltages, $V \lesssim 0.01\Delta/e$, the direct conductances vanish, while the transconductances become quantized, $G_{\alpha \beta} = -(4e^2/h) C_{\alpha \beta}$. Namely, for $\phi_0=0$ (trivial region), $G_{12}=-G_{21}=0$ and, for $\phi_0=2.21$ (topological region), $G_{12}=-G_{21}=4e^2/h$.}

The Landau-Zener transition rate at an avoided crossing between the two states is given as $\Gamma_{\rm LZ}=peV/\pi$ with $p=\exp[-\pi E_g^2/v]$. Here $E_g=E(t^*)$ and $v=\partial E/\partial t(t\gg t^*)$ for $E(t)=E_1(\phi_0, 2n_1 eVt,2n_2eVt+\phi)$ having an avoided crossing at $t^*$. 
From the central panel of Fig.~\ref{fig:ABS}, which shows the cut in the plane of phases $(\phi_1,\phi_2)$ going through the minimal gap, we extract $E_{g} = 0.11\Delta$ and $v \sim eV \Delta$ for the scattering matrix $\hat{S}_1$. Using these values, our estimate for the Landau-Zener transition rate becomes $\Gamma_{\rm LZ} \approx \Gamma = 0.002 $ at $V \approx 0.02 \Delta / e$, which is in good agreement with the voltage where one starts to see low dissipation and the quantization of the transconductance (cf.~Figs.~\ref{fig:conductance} and \ref{fig:dissipation}). 

{ When approaching the Weyl points, the gap in the $(\phi_1,\phi_2)$-plane decreases. Thus, the voltage $V^*$ below which conductance quantization can be observed decreases as well. As shown in Fig.~\ref{fig:G-phi0}, at  fixed voltage, we see a peak in the direct conductances around the Weyl points signaling that dissipation is large (see also Fig.~\ref{fig:dissipation-phi0} in Appendix \ref{app-B}). The smaller the voltage, the more one can approach the Weyl points without losing the transconductance quantization.}

\section{Conclusion}
\label{sec:5}

It has recently been predicted that multiterminal Josephson junctions may realize a novel type of topological matter.\cite{Riwar2016} Namely, for $n \geq 4$ terminals, Weyl singularities may appear in the Andreev bound state spectrum of the junction, giving rise to topological transitions as the superconducting phases are tuned. These transitions are observable as quantized jumps in the transconductance between two voltage-biased terminals. In this work, we have studied this effect by numerically solving the Landauer-B\"uttiker scattering theory for a 4-terminal Josephson junction, which describes the quasiparticle transfer between the leads by the process of multiple Andreev reflection in the subgap regime. We have observed how the transconductances approach the quantized values predicted by the topology at low voltages, when dissipation is small. Tuning the superconducting phase at a fixed voltage, the topological transitions could be clearly seen. Our results provide an important step towards the clarification of the experimental conditions to observe the topological properties of multiterminal Josephson junctions.

\acknowledgments
This work was supported by ANR, through grant ANR-12-BS04-0016-03, and by the Nanosciences Foundation in Grenoble, in the frame of its Chair of Excellence program.

\begin{widetext}

\appendix
\label{appendix}

{ \section{Statistics of the largest possible spectral gaps for random scattering matrices with Weyl points}
\label{app-A}

In order to find suitable scattering matrices for our numerical investigation, we analyzed an ensemble of random scattering matrices. In particular, we searched for the largest possible gap in the $(\phi_1, \phi_2)$-plane, both in the topologically trivial and the topologically non-trivial regime. A histogram of the smaller of these maximal gaps in the two regions for an ensemble of topological scattering matrices is shown in Fig.~\ref{fig:histogram}. For our simulation, we choose two different topological scattering matrices: matrix $\hat S_1$, for which results are presented in the main text, with a gap close to the largest value we could obtain and matrix $\hat S_2$, for which results are presented in Appendix \ref{app-C}, with a more typical gap.

\begin{figure}
	\centering
		\includegraphics[width=0.4\textwidth]{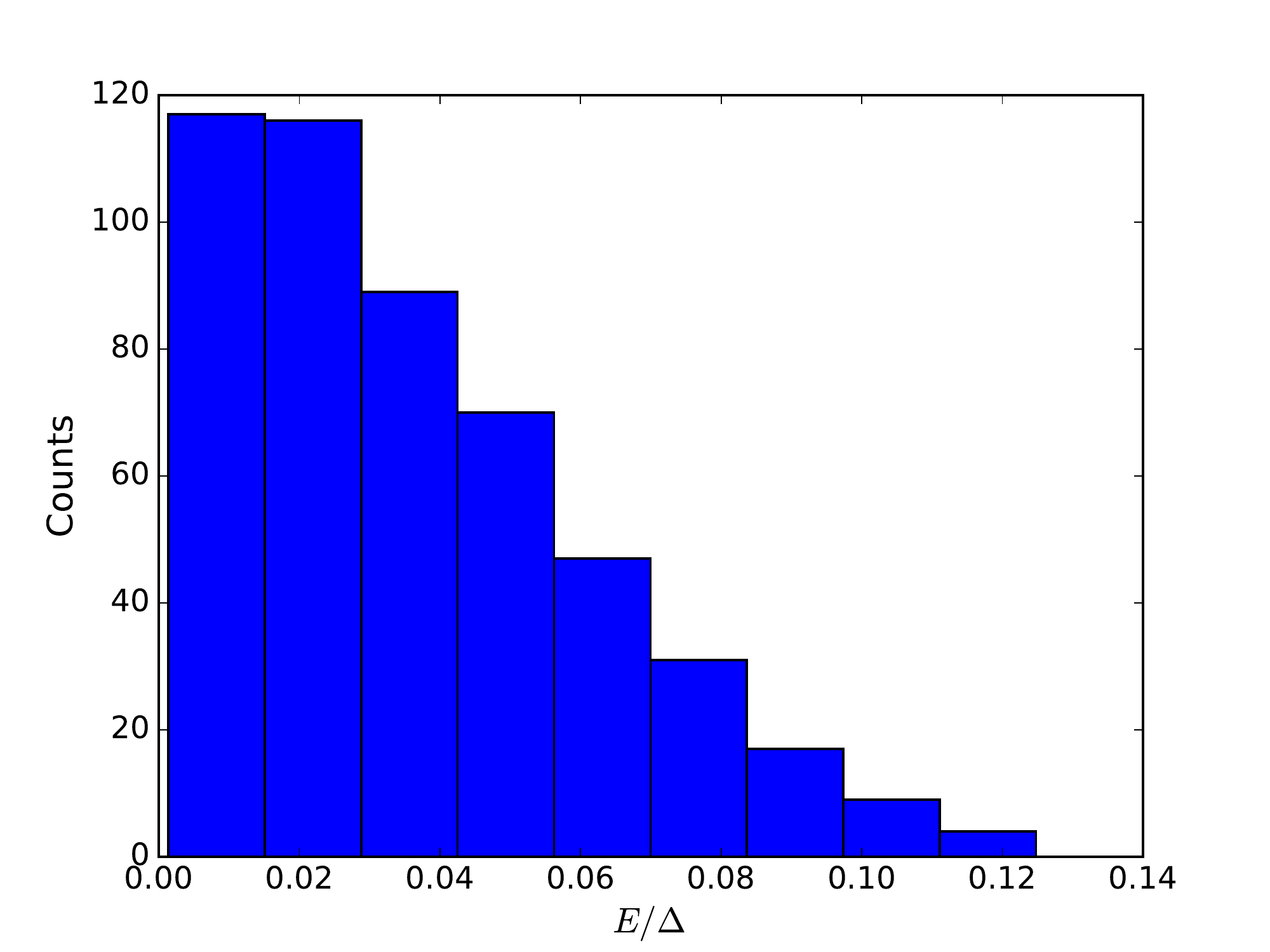}
	\caption{To observe the quantization of the transconductance, it is favorable to have $(\phi_1, \phi_2)$-planes with a large spectral gap both in the topological and the trivial region. Here we show a histogram of the smaller of the maximal gaps in these two regions for $N=500$ random scattering matrices with Weyl points.}
	\label{fig:histogram}
\end{figure}}

\section{Additional results for the scattering matrix $\hat S_1$}
\label{app-B}

To obtain the results presented in the main part, we averaged the currents over the phase offset $\phi$ between the phases of the leads 1 and 2. As shown in Fig.~\ref{fig:offset}, the dependence of the currents on $\phi$ is weak and smooth, justifying this procedure. 

 \begin{figure}[h!]
	\centering
		\includegraphics[width=0.35\textwidth]{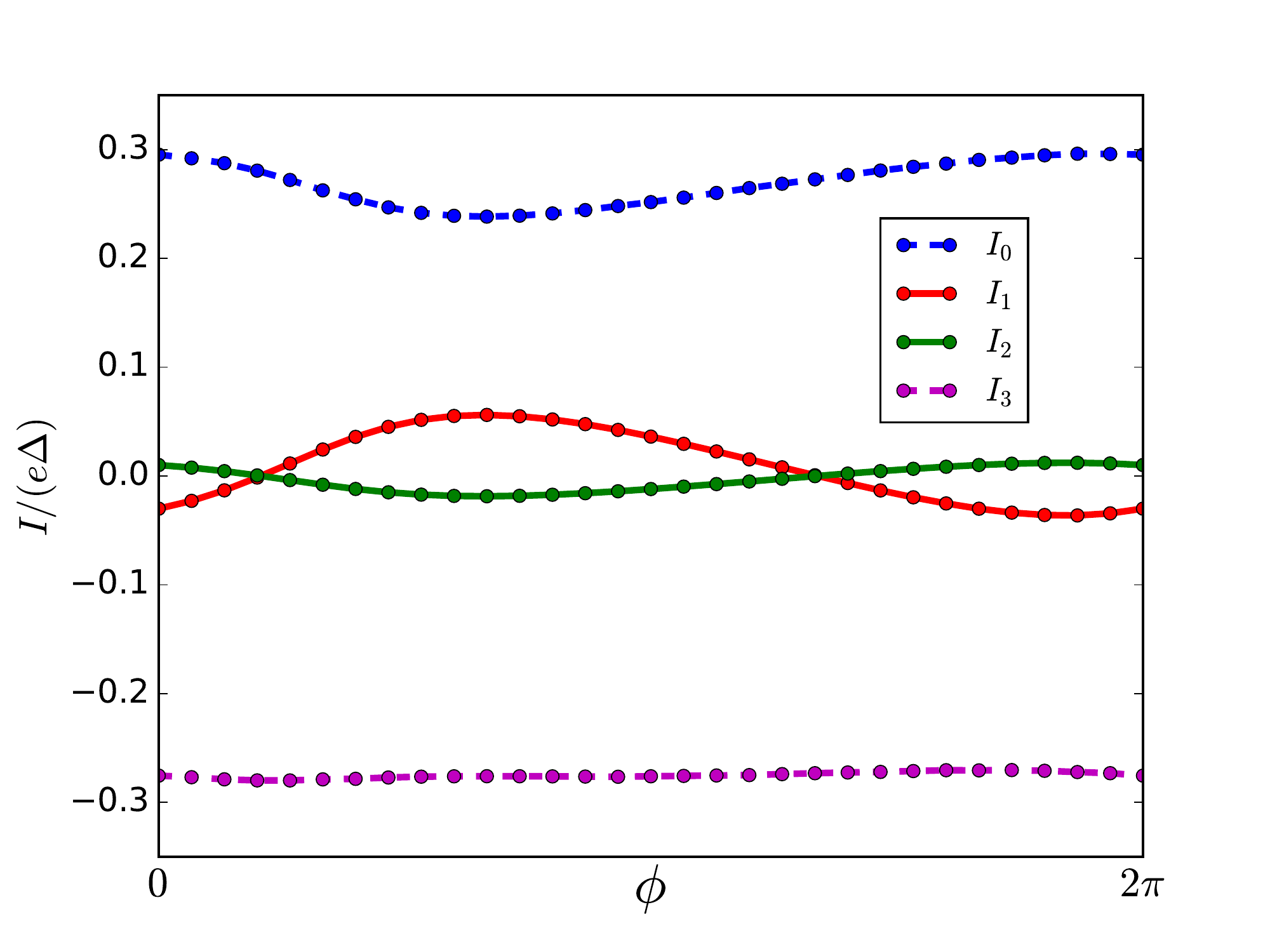}
		\includegraphics[width=0.35\textwidth]{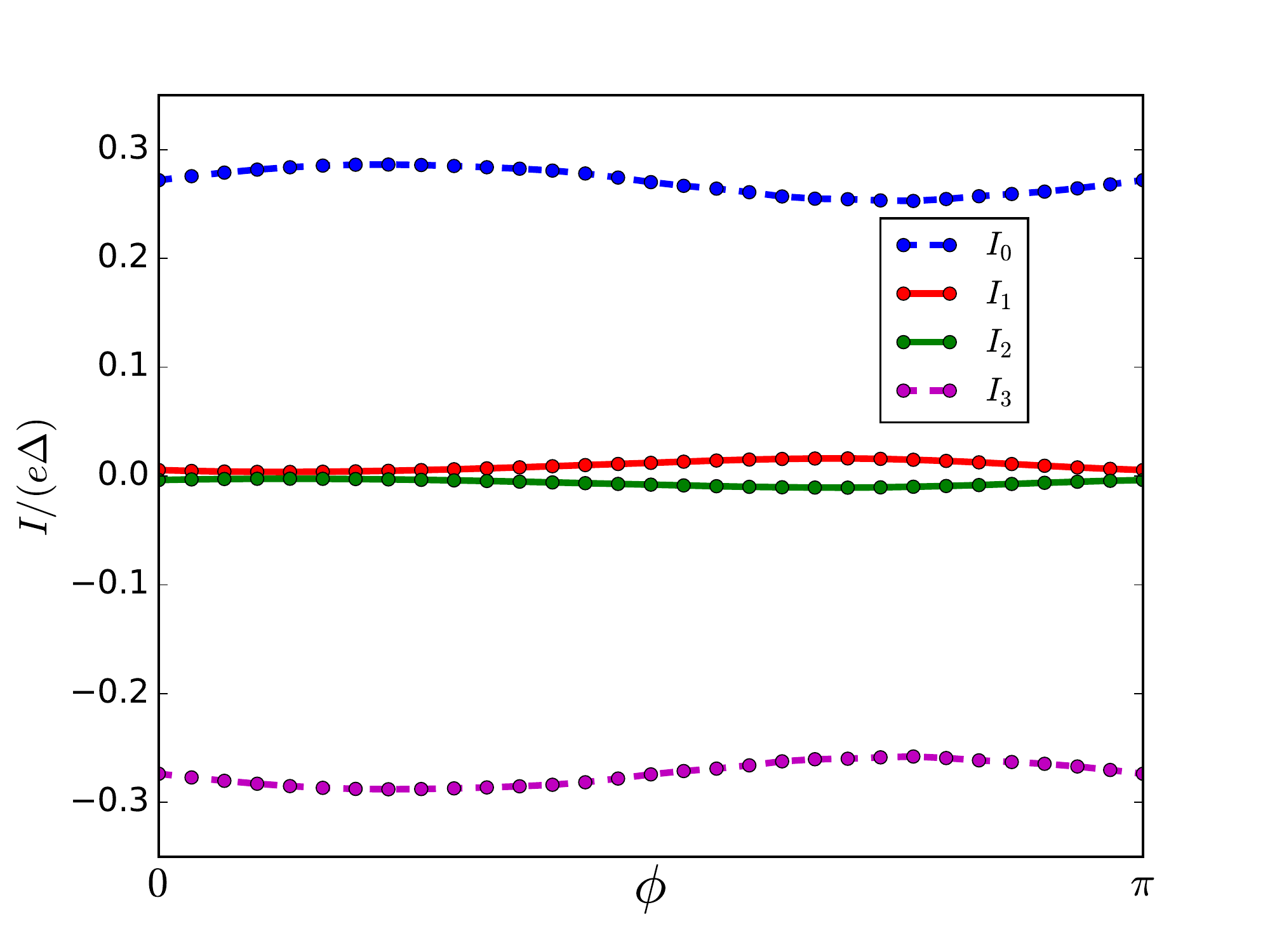}
	\caption{Dependence of the currents on the phase shift $\phi$. Here $\phi_0=2.21$, $\Gamma=0.002\Delta$ and $V=0.005\Delta/e$.  Left: $\phi_1=eVt,\phi_2=3eVt+\phi$. Right: $\phi_1=2eVt,\phi_2=3eVt+\phi$. We see that already $N=10$ phase shifts give a rather good sampling.}
	\label{fig:offset}
\end{figure}

To observe the quantization of the transconductance, transport has to be quasi-adiabatic, i.e., dissipation has to be low. Close to the Weyl points, the gap around the Fermi level becomes very small and this breaks down. The dissipation as a function of the control parameter $\phi_0$ is shown in Fig.~\ref{fig:dissipation-phi0}. Large peaks at the positions of the Weyl points are clearly visible. We also show the chirality $C$ that is expected to be zero in the trivial region and $\pm1$ in the topological region. Due to the large dissipation, it deviates from these values in the vicinity of the Weyl points.

\begin{figure}[h!]
	\centering
		\includegraphics[width=0.47\textwidth]{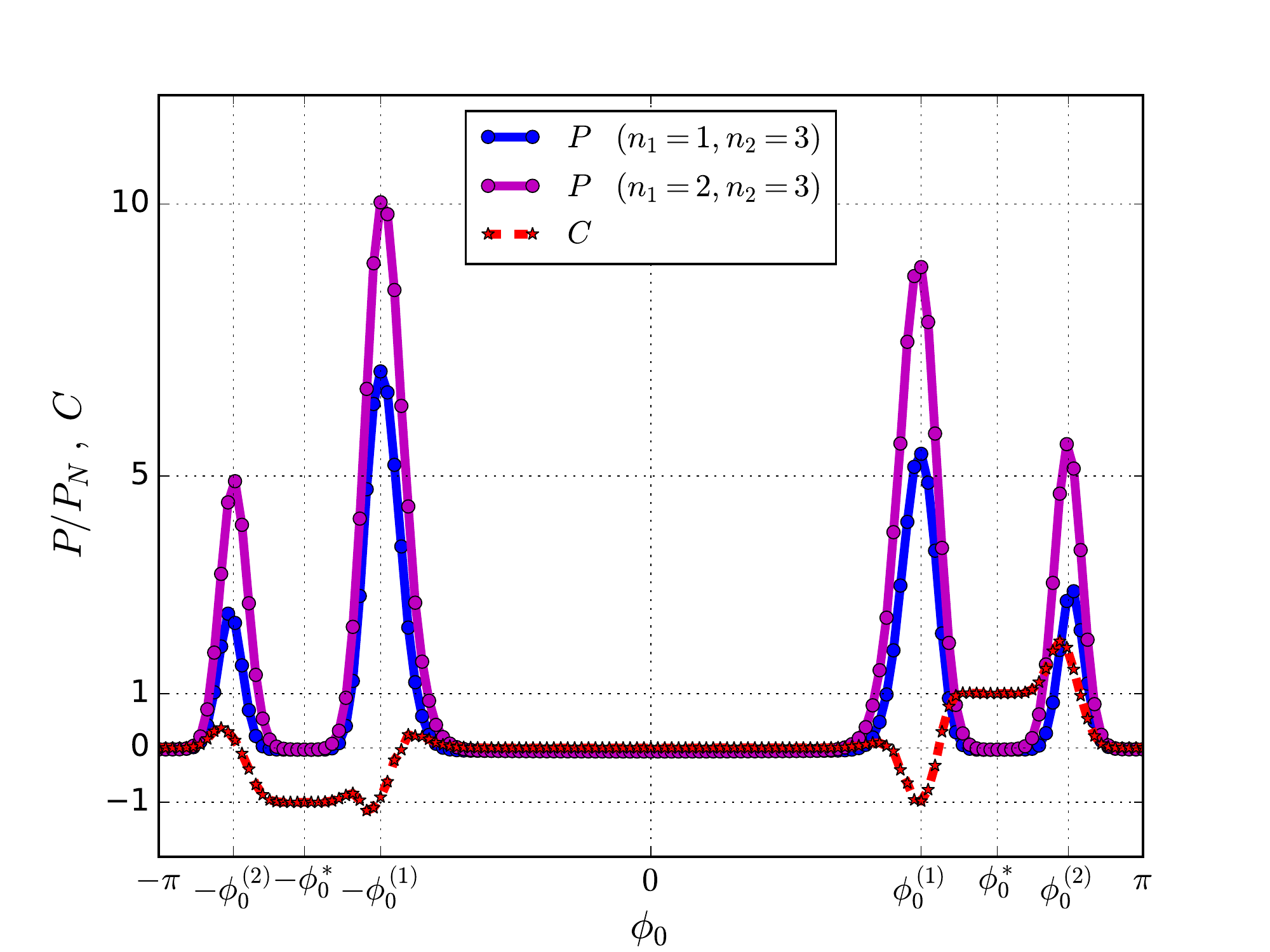}
	\caption{Dissipation and chirality  as a function of phase $\phi_0$ at fixed voltage  $V=0.003 \Delta/e$. The two curves for dissipation correspond to two sets of voltages $V_1=n_1V,V_2=n_2V$.}
	\label{fig:dissipation-phi0}
\end{figure}

\section{Scattering matrix $\hat S_2$}
\label{app-C}

The results presented in the main text were obtained for a scattering matrix that yields a particularly large gap in the topological region. As can be seen from Fig.~\ref{fig:histogram}, typical scattering matrices yield a smaller gap.  In this section we make use of a second scattering matrix that is more typical,
\begin{equation} \label{S2}
\hat{S}_2 = \left(  \begin{matrix}
 0.108-i0.144&  0.180-i0.119 & 0.185-i0.590& 0.734+i0.015\\
0.180-i0.119 & 0.151+i0.234& -0.362-i0.634& -0.4750+i0.341\\
0.1852-i0.590 &-0.362-i0.634 & 0.145-i0.009 & -0.204+i0.146\\
0.734+i0.015& -0.475+i0.341 &-0.204+i0.146& 0.236-i0.022
\end{matrix}  \right).
\end{equation}
Here the Weyl points are at $\pm (1.74,-1.07,-2.82)$, with charge $+1$, and $\pm (2.50,3.02,-0.73)$, with charge $-1$. The gap in the topological region is largest in the planes at $\phi_0 =\pm 2.16$, where $E = 0.05 \Delta$.

Due to the smaller gap, lower voltages have to be used to observe the quantization of the transconductance, making the calculations much more time-consuming. The current-voltage characteristics are shown in Fig.~\ref{fig:currents2}. The conductances are shown in Fig.~\ref{fig:conductance2}. { As explained in the main text, four different voltage regimes can be distinguished. At high voltages, $V\gg\Delta/e$, one find the normal state conductances, $G^N_{11}=-0.46\times(4e^2/h)$, $G^N_{22}=-0.49\times(4e^2/h)$, and $G^N_{12}=G^N_{21}=0.27\times(4e^2/h)$. Lowering the voltage, $0.01\Delta/e\lesssim V\lesssim 2\Delta/e$, multiple Andreev reflections lead to resonance features. At even lower voltages, $V\lesssim 0.01\Delta/e$, the interplay between Landau-Zener transitions and inelastic relaxation becomes important. Here we chose a Dynes parameter $\Gamma = 0.0001\Delta$. At $\phi_0 = 2.16$, it becomes comparable to the Landau-Zener rate at voltage $V\approx 0.002 \Delta/e $. This is consistent with the observed quantization at $V\lesssim 0.001\Delta/e$. The dissipation and the chirality are shown in Fig.~\ref{fig:dissipation2}. }

  \begin{figure}[h!]
	\centering
		\includegraphics[width=0.45\textwidth]{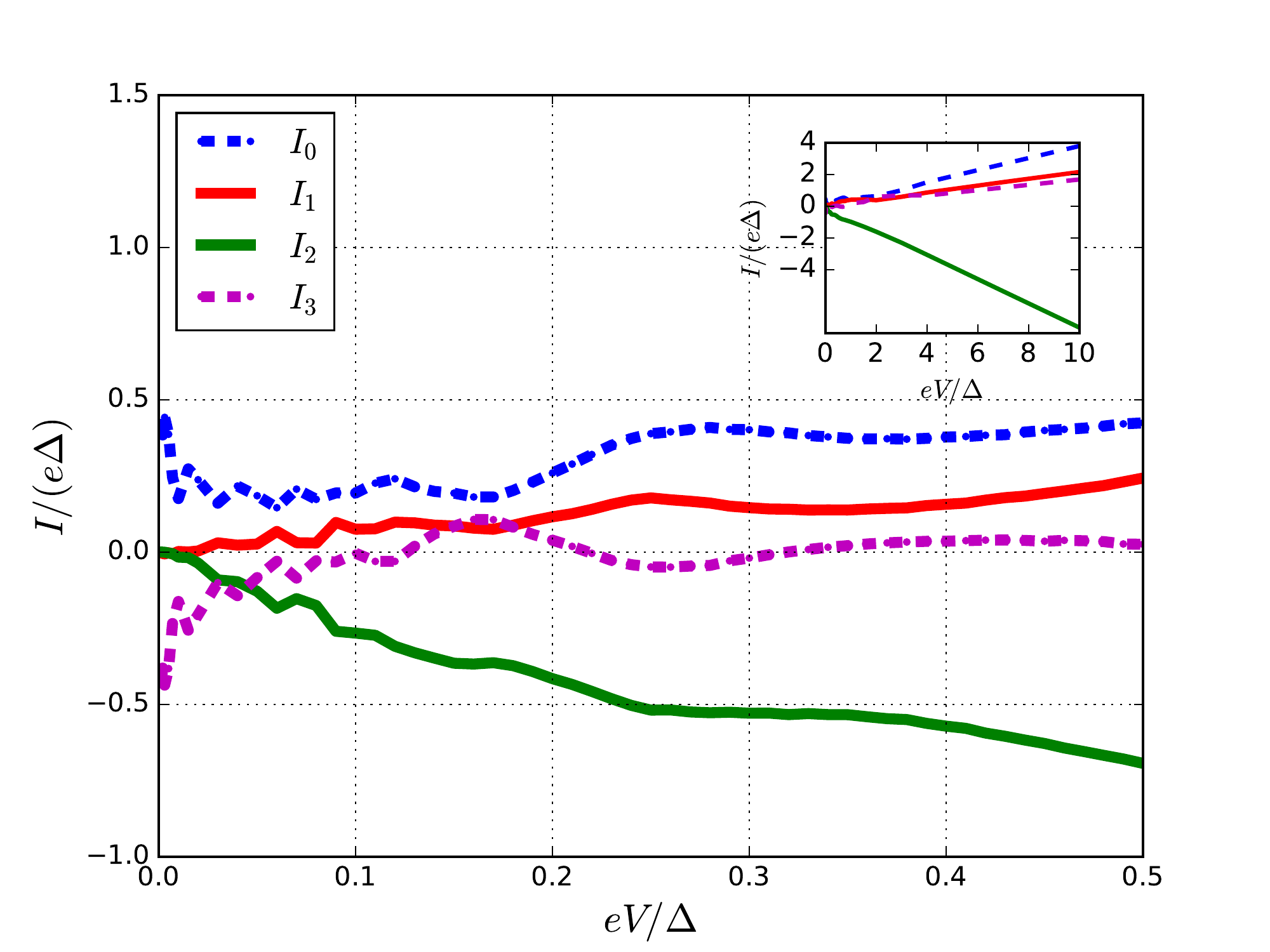}
	    \includegraphics[width=0.45\textwidth]{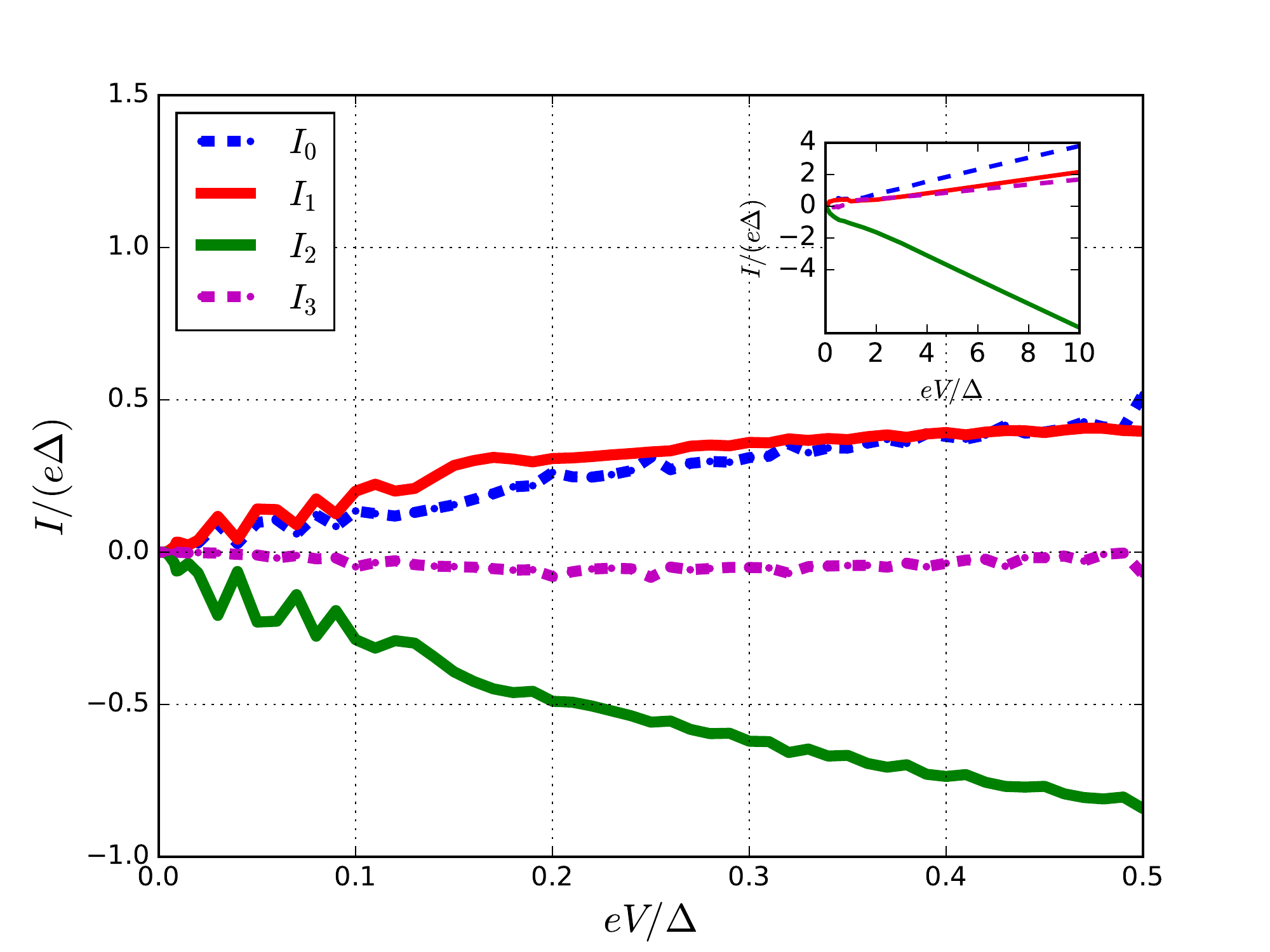}\\
	    \includegraphics[width=0.45\textwidth]{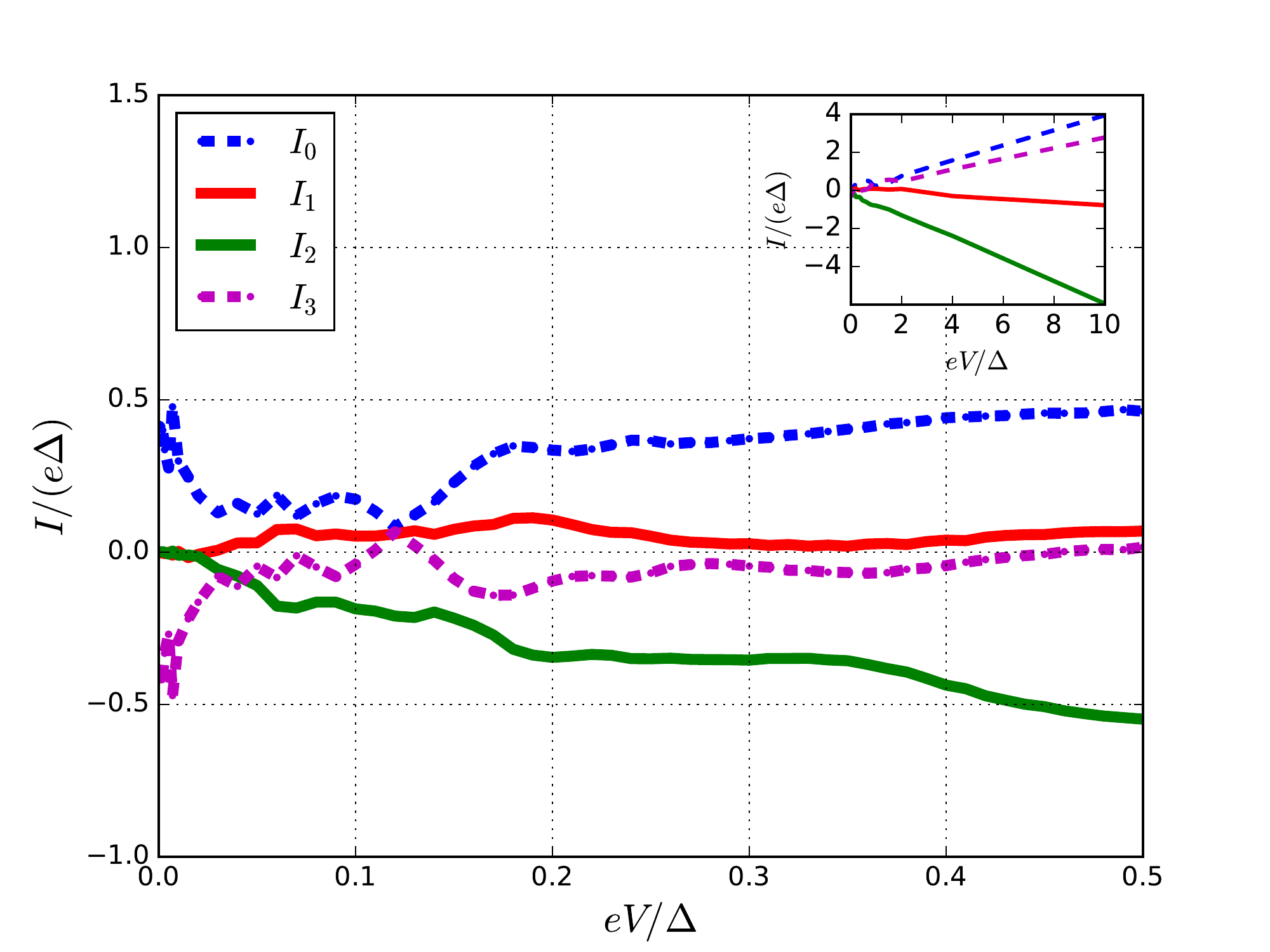}
	    \includegraphics[width=0.45\textwidth]{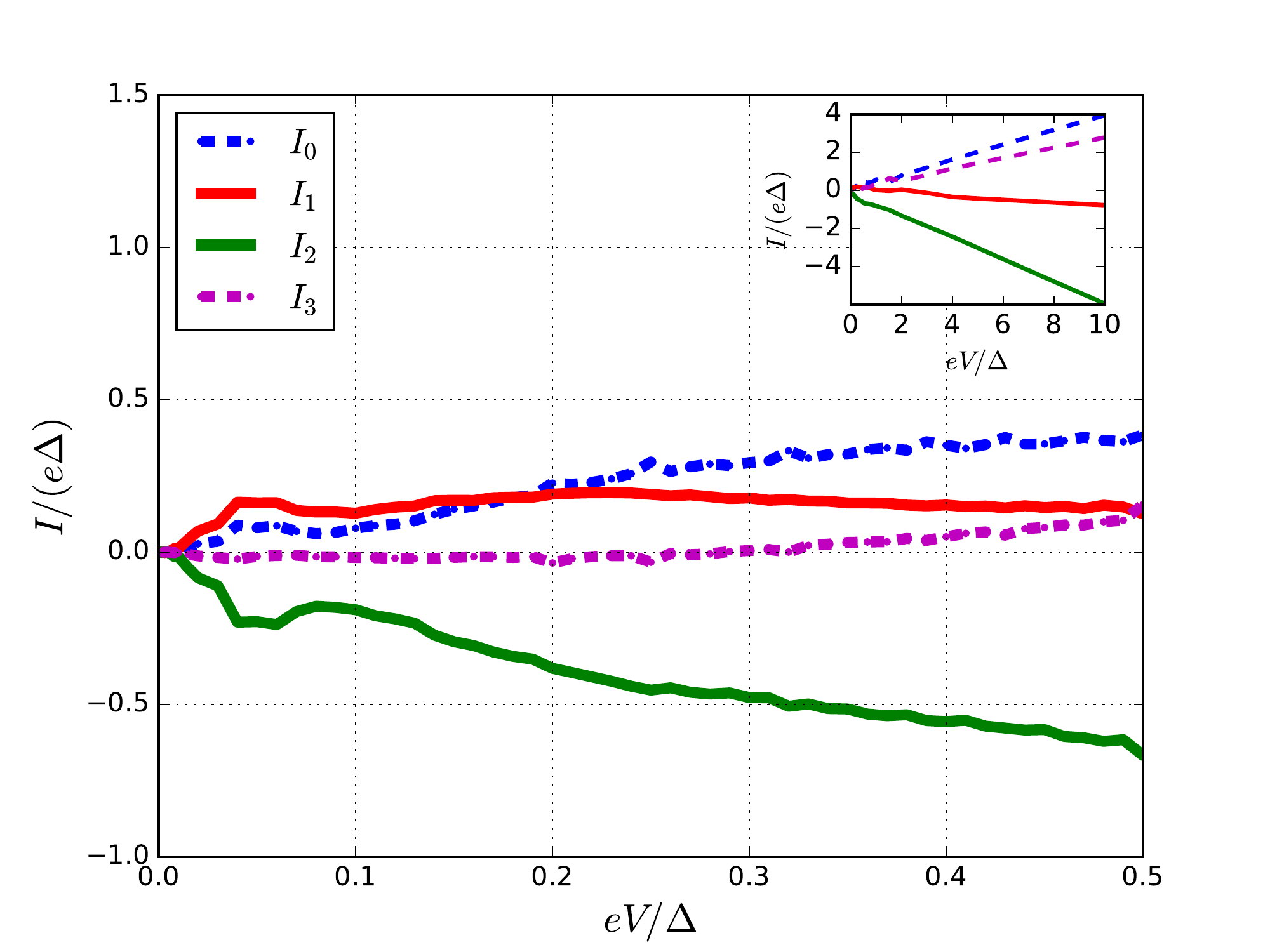}
	\caption{The currents $I_0,I_1,I_2,I_3$ as function of voltage for the scattering matrix $\hat{S}_2$. The voltages in terminals $1$ and $2$ are given as $V_1 = n_1V$ and $V_2 =n_2V$, respectively (top: $n_1=1,n_2=3$, bottom: $n_1=2,n_2=3$). The Dynes parameter is set to $\Gamma=0.0001\Delta$. Left: At phase $\phi_0 = 2.16$ in the topological region. Right: At phase $\phi_0=0$ in the trivial region. We have used an average over $N=10$ phase shifts $\phi$. The insets show a larger range of voltages.}
	\label{fig:currents2}
	\end{figure}

  \begin{figure}[h!]
	\centering
		\includegraphics[width=0.49\textwidth]{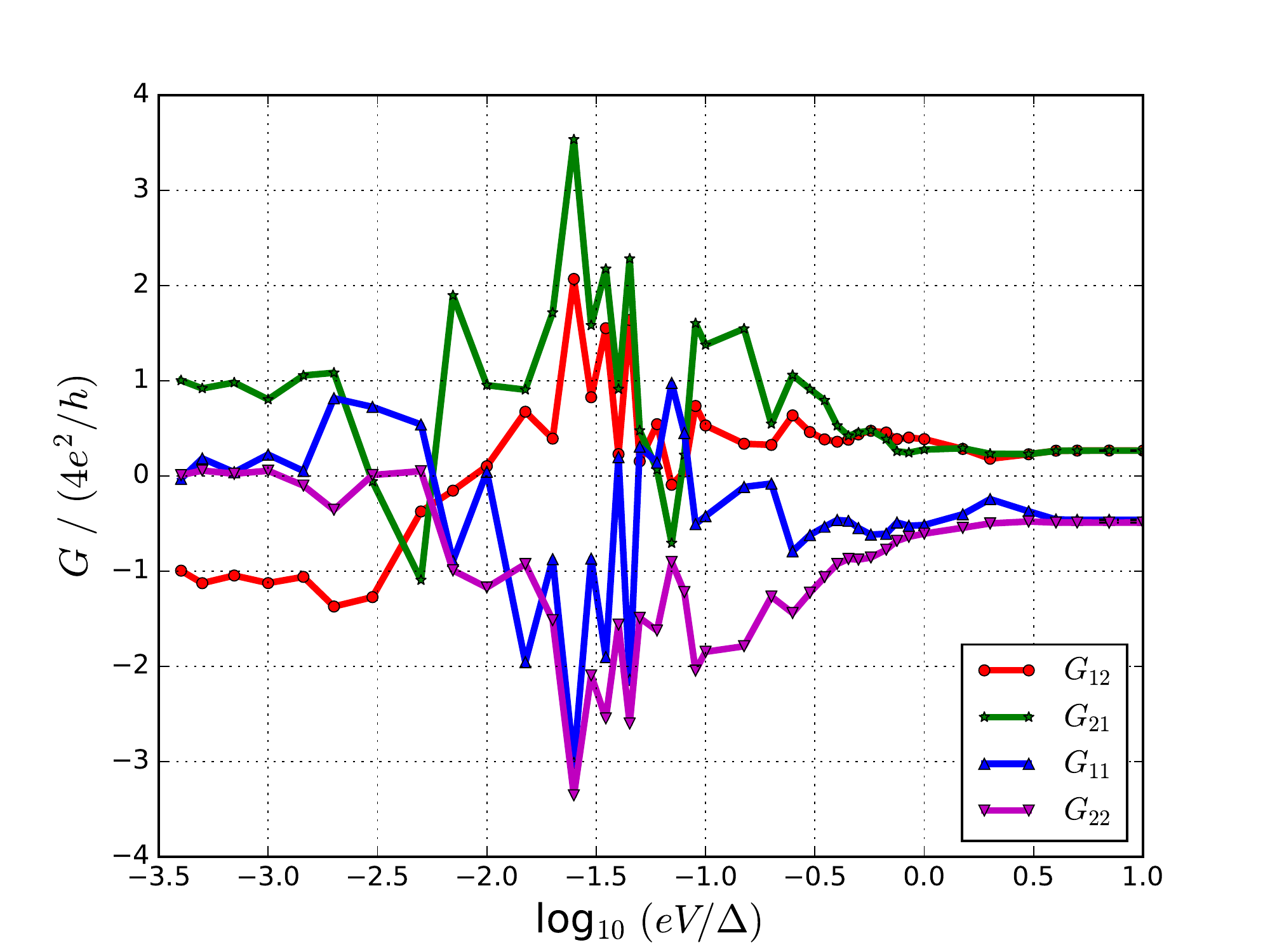}
		\includegraphics[width=0.49\textwidth]{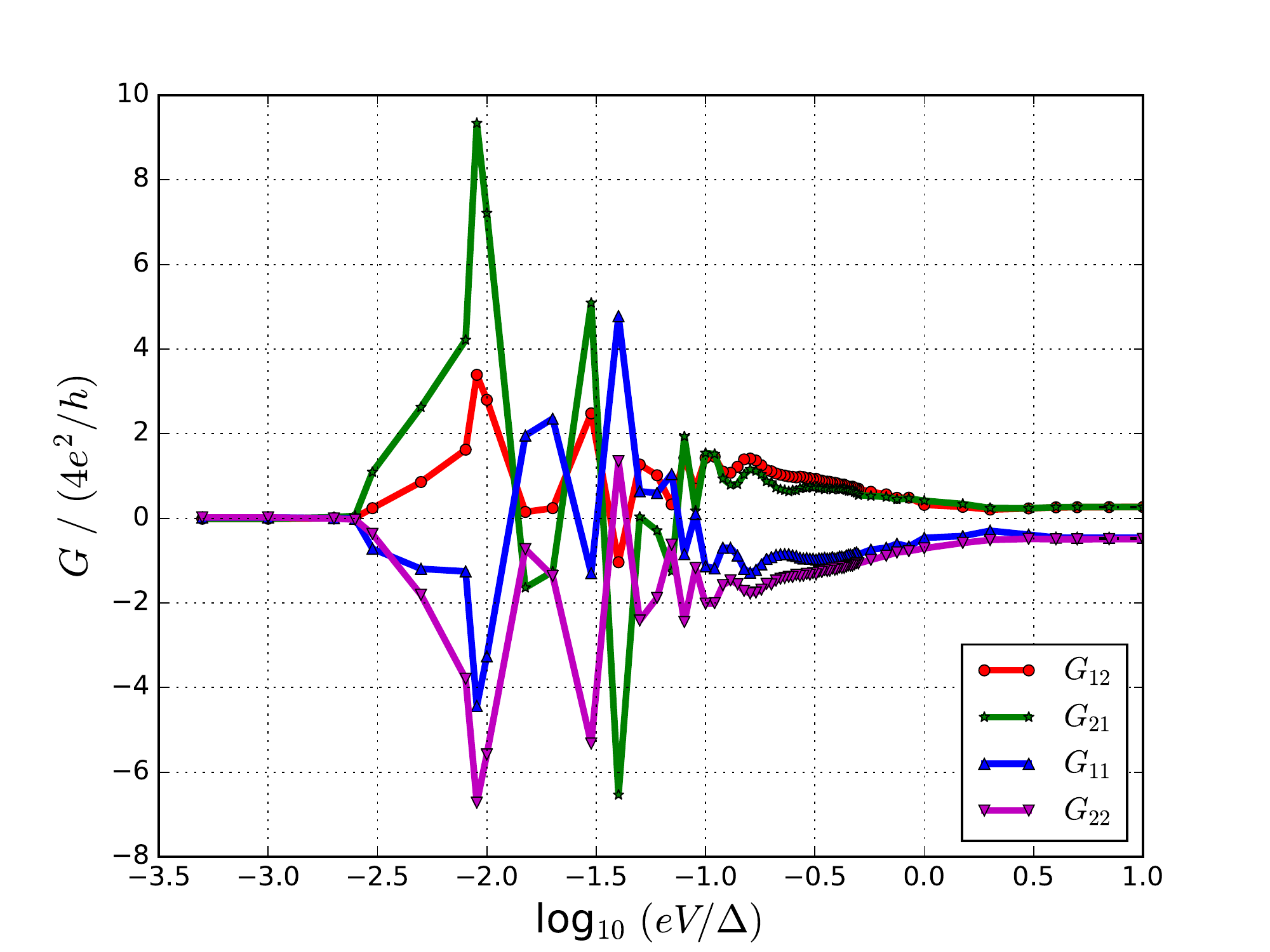}
	\caption{
	The conductances $G_{12},G_{21},G_{11}$, and $G_{22}$ between the voltage-biased leads $1,2$ as a function of voltage in logarithmic scale. The conductances are obtained from the currents shown in Fig.~\ref{fig:currents2} . Left: At phase $\phi_0 = 2.16$ in the topological region. Right: At phase $\phi_0=0$ in the trivial region. The expected quantization of the transconductance is seen for voltages $eV/\Delta\lesssim 0.001$.}
	\label{fig:conductance2}
\end{figure}
	
 \begin{figure}[h!]
	\centering
		\includegraphics[width=0.35\textwidth]{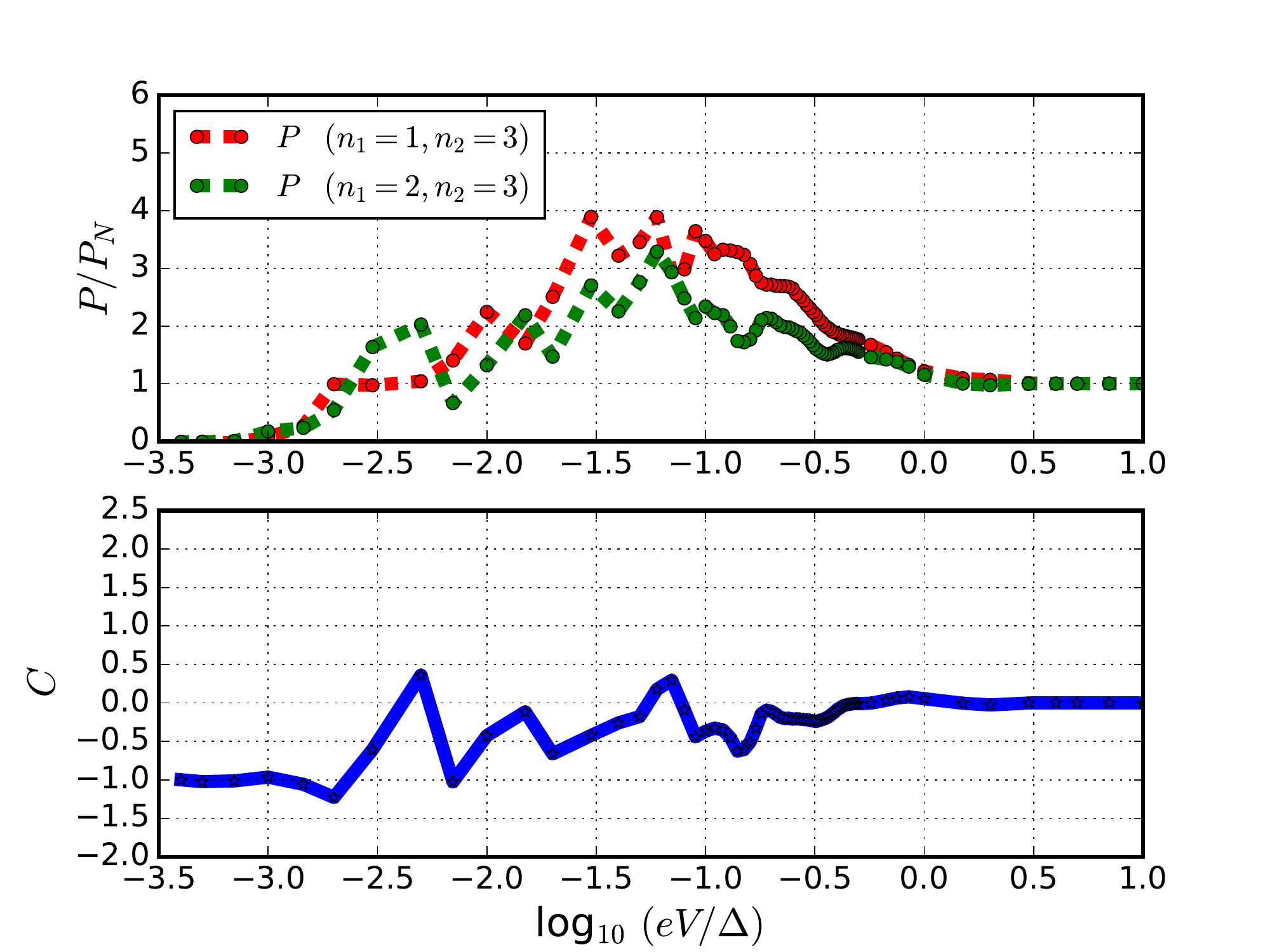}
		\includegraphics[width=0.35\textwidth]{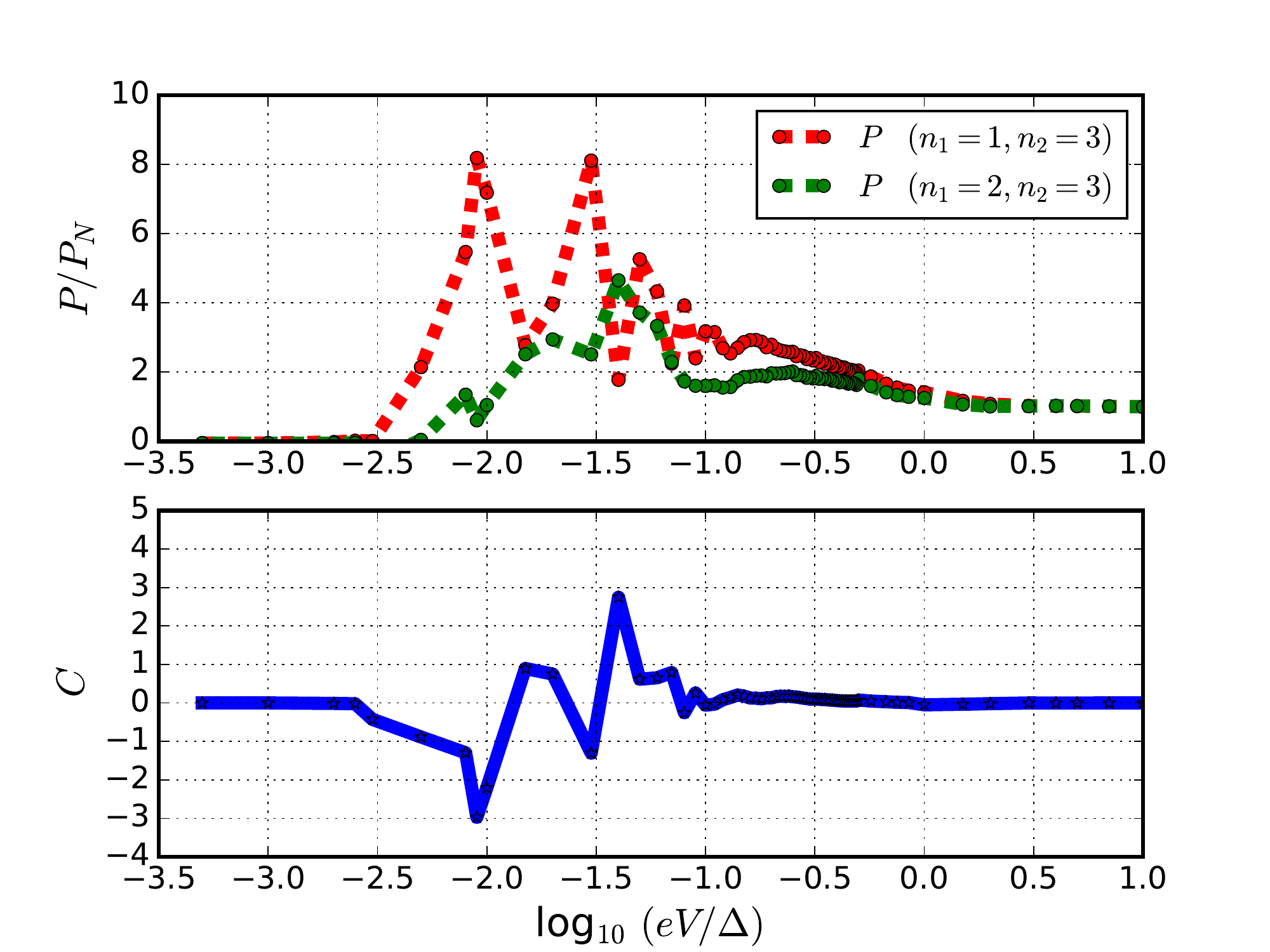}
	\caption{Dissipation $P/P_N$ (dashed lines) and chirality $C$ (solid lines) as a function of voltage in logarithmic scale. Green curves correspond to $V_1=2V,V_2=3V$ and red curves to$V_1=V,V_2=3V$.  Left: At phase $\phi_0 = 2.16$ in the topological region. The chirality tends to $-1$ as the dissipation tends to zero at low enough voltages. Right: At phase $\phi_0=0$ in the trivial region.}
	\label{fig:dissipation2}
\end{figure}

\end{widetext}

\end{document}